\documentclass[journal=aelccp,manuscript=article]{achemso}
\setkeys{acs}{articletitle = true}

\usepackage{graphicx}
\usepackage{dcolumn}
\usepackage{bm}

\usepackage{achemso}
\usepackage{amsmath}
\usepackage{mhchem}
\usepackage{color,soul}
\soulregister\cite7
\soulregister\ref7

\usepackage{tabu}
\usepackage{nth}
\usepackage[dvipsnames]{xcolor}
\definecolor{orange}{rgb}{1.0,0.4,0.0}
\usepackage{MnSymbol}
\usepackage{framed}

\newcommand{\ket}[1]{\left| #1 \right\rangle}

\newcommand{\abs}[1]{\left| #1 \right|}

\newcommand{\textapprox}{{\raise.17ex\hbox{$\scriptstyle\mathtt{\sim}$}}}

\usepackage{accents}

\usepackage{caption}

\title{The effects of decoherence on Fermi's golden rule}

\author{Caihong Zheng}
 \affiliation{School of Physical Science and Technology, ShanghaiTech University, Shanghai 201210, China. }

 \author{Fan Zheng}
 \affiliation{School of Physical Science and Technology, ShanghaiTech University, Shanghai 201210, China. }
 \email{zhengfan@shanghaitech.edu.cn}

\begin{document}
\maketitle

\captionsetup[figure]{name={Fig.}}

\begin{abstract}
Fermi's golden rule which describes the transition rates between two electronic levels under external stimulations is used ubiquitously in different fields of physics. The original Fermi's golden rule was derived from perturbative time-dependent Schr\"{o}dinger's equation without the direct contribution by decoherence effect. However, as a result of recent developments of quantum computing and ultra fast carrier dynamics, the decoherence becomes a prominent topic in fundamental research.Here, by using the non-adiabatic molecular dynamics which goes beyond the time-dependent Schr\"{o}dinger's equation by introducing decoherence, we study the effect of decoherence on Fermi's golden rule for the fixed basis and the adiabatic basis, respectively. We find that when the decoherence time becomes short, there is a significant deviation from the Fermi's golden rule for both bases. By using monolayer $\mathrm{WS_2}$ as an example, we investigate the decoherence effect in the carrier transitions induced by the electron-phonon coupling with first-principle method.
\end{abstract}

Fermi's golden rule (FGR) is used in almost every subfield in physics~\cite{Sakurai17p}. It describes the transition from one electronic state to another under an external stimulation. If two states have energy $\epsilon_1$ and $\epsilon_2$, respectively, and the stimulation has a frequency $\omega$, then an energy conservation rule $\delta(\epsilon_1-\epsilon_2-\hbar\omega)$ must be satisfied. This is derived from a perturbative time-dependent Schr\"{o}dinger's equation (TDSE)~\cite{Sakurai17p}. Many other effects, e.g. the phonon assisted transition, or other random stimulus not included in the primary frequency $\omega$ stimulus, are often represented by a Lorentzian broadening of the delta function. If the stimulation itself has a finite decay time, then its Fourier spectrum does not have a single-frequency $\omega$, which also introduces a Lorentzian broadening lineshape on the delta function, e.g.: $(1/\gamma)/\left[\left(\epsilon_1-\epsilon_2-\hbar\omega\right)^2+(1/\gamma)^2\right]$. Here $\gamma$ is the exponential decay time of the stimulation.
All these, however, do not include the decoherence (or dephasing) between the states 1 and 2 themselves. Under quantum mechanics, the phases of states 1 and 2 are changing according to $e^{-i\epsilon_1 t}$ and $e^{-i\epsilon_2 t}$, respectively. The decoherence means that in addition to the above phase changes, the overlaps between these two states will reduce with time exponentially as $e^{-t/\tau}$ where $\tau$ is usually called the decoherence time. This reduction is often caused by additional components in the wavefunction not represented by the primary electronic state description (see discussion in Ref.2). One example is the phonon degrees of freedom in a system, with their dynamics depending on their associated electronic states. As a result, the phonon dynamics will be different for a system in electronic states 1 and 2. This leads to a deviation of their trajectories, and an exponential decay of their phonon wavefunction overlaps. Besides phonon, other degree of freedom like the nuclear spin can also cause decoherence~\cite{Seo16p12935a,Graham17p3196}.
Such decoherence phenomena are prominent in quantum computing as well as ultrafast electron dynamics (hot carrier cooling~\cite{Bernardi15p5291,Jhalani17p5012,Bernardi14p257402,Jalabert90p3651,Jacoboni83p645,Sjakste07p236405,Giustino17p015003} and laser induced phase transition~\cite{Paillard19p087601,Liu20p184308} etc).  One fundamental question is: how dose the decoherence effect influence the FGR. In this study, using the external stimulus-induced excitations in an electronic system as an example, the role of decoherence to the carrier excitation is explored. We find that when $\tau$ is becoming small (i.e. the decoherence effect is becoming strong), a dramatically different formalism is required. Depending on different basis, the role of decoherence effect will also be different. We use non-adiabatic molecular dynamics (NAMD) to include the decoherence effects in the description of the electron dynamics. The NAMD simulation is also performed on a real physical system (monolayer $\mathrm{WS_2}$) by using first-principle calculations. The influence of decoherence on transition in this system is consistent to our prediction.

Decoherence process is an important element in various NAMD schemes \cite{Zheng17p6435,Long16p1996,Fischer09p15483,Zhang18p6057,Zhang18p1112,Ren13p205117,Kang19p224303,Zheng19p6174}. Originated from the full quantum treatment of electrons and ions' motion, decoherence describes that when the wavefunctions of ions are rapidly diverging from each other, their corresponding electronic wavefunctions are less likely to interfere with each other and the built-up wavefunction-mixing by the non-adiabatic dynamics tends to ``collapse". Decoherence has been incorporated in various NAMD schemes with the mixed quantum-classical approach. These include wavefunction coefficient corrections for fewest-switches surface hopping (FSSH)\cite{Tully90p1061,Granuccip10,Soudackov11p144115,Subotnik11p024105a}, wavefunction collapsing\cite{Jaeger12p22A545,Wang20p9075}, decaying coherence in Liouville–von Neumann equation\cite{Wong02p8429}, and the P-matrix method~\cite{Kang19p224303,Zheng19p6174}. In particular,  by employing the classical path approximation, NAMD as a post-processing of \textit{ab initio} molecular dynamics (MD), allows a direct simulation of excited carrier movements with almost negligible cost of additional calculations\cite{Akimov13p4959,Akimov14p789,Zheng17p6435,Zhang19p6151,Kang19p224303,Zheng19p6174}.  Although both NAMD and FGR are widely used to describe the electronic state transitions, surprisingly, there are not many works comparing these two approaches directly. Thus, it will be extremely useful to reveal how decoherence time influences the transition rate and how this is compared to FGR calculation.

One of the challenges of comparing FGR and NAMD stems from the fact that even the simplest system such as bulk Si or CO$_2$ molecule experiences complicated dynamical trajectories at a finite temperature: such as band crossing and higher order electron-phonon couplings (EPC). Therefore, in this work, a two-level model system (TLS) is used as an example to compute the carrier excitation with both FGR and NAMD.  Albeit its simplicity, TLS plays an important role in various fundamental studies. For example, TLS as a qubit is the most basic component in quantum computing\cite{Graaf20peabc5055,KlimovP.V.2018FoET}. 

The Hamiltonian for the TLS under an external stimulation can be written as: 
\begin{align}
    H = \begin{bmatrix}
         \epsilon_1  &  V_0e^{i\omega t} \\
         V_0 e^{-i\omega t} & \epsilon_2
    \end{bmatrix}\label{eq1}
\end{align}

\noindent where states $\ket{1}$ and $\ket{2}$ with corresponding energies $\epsilon_1$ and $\epsilon_2$ respectively are coupled with a sinusoidal-type external field with amplitude $V_0$ ($V_0$ sets to be real). In reality, the external field could be any time-evolving single-frequency $\omega$ field which couples states $\ket{1}$ and $\ket{2}$. The adiabatic states (or time-dependent eigen states) obtained by diagonalizing the Hamiltonian are labeled as $\ket{\phi_1(t)}$ and $\ket{\phi_2(t)}$. $E_i(t)$ ($i=1,2$) are their eigen energies. The time-dependent wavefunction evolution describing the electron can be generally written as $\ket{\psi}=\sum_{i=1,2}c_i\ket{g_i}$, which follows the Schr\"{o}dinger's equation. $c_i$ are the coefficients on the basis $\ket{g_i}$. Since the decoherence effect is not a unitary transformation, using different basis will make a difference. In this work, we have chosen two widely used basis: i) fixed basis (i.e. $\ket{1}\ \mathrm{and}\ \ket{2}$ as the basis and they are invariant during the evolution), ii) adiabatic basis (i.e. $\ket{\phi_1} \ \mathrm{and}\ \ket{\phi_2}$ are eigenstates of the time-dependent Hamiltonian). 
Here, $V_0$ is set to be sufficiently small compared to $\hbar\omega_{12}$ ($\hbar\omega_{12}=\left|\epsilon_1-\epsilon_2\right|$) to avoid higher-order coupling in NAMD simulation.
Equivalently, the density matrix $D_{ij}(t)$ can be used to describe the evolution. For the fixed basis, $D_{ij}(t)$ could be solved easily by numerically integrating the Schr\"{o}dinger's equation (SI equ.3). However, for the adiabatic basis, an efficient evolution for $D_{ij}(t)$ or the wavefunction is not trivial, and we choose the so-called P-matrix method\cite{Kang19p224303,Zheng19p6174} to compute $D_{ij}(t)$ with the decoherence effect included (SI equ.11). Compared to other NAMD implementations such as FSSH and decoherence induced surface hopping (DISH), P-matrix implements detailed balance and decoherence together in one simulation. More importantly, based on the density matrix evolution, P-matrix only requires a single-shot calculation to yield the statistical assembly, which circumvents the need for statistical averaging over thousands of trajectories used in other stochastic based NAMD methods.

As discussed before, FGR has a delta function to preserve the energy conservation ($\hbar\omega = \hbar\omega_{12}$). However, such delta function brings difficulties to compare with NAMD numerically. In practice, a broadened Lorentzian is usually applied to replace the delta function. This broadening can come from additional external effects or from the damping of the stimulus itself. We introduce a damping for the external perturbation: the off-diagonal element of the Hamiltonian (equ.1) is $V_0e^{i\omega t}e^{-\eta t}$ and $\eta$ is a small positive number. Following the perturbation theory derivation, if the initial condition has state $\ket{1}$ fully occupied at $t=0$, the analytical solution of the TDSE yields the state $\ket{2}$ occupation (projected the electron wavefunction as $\ket{\psi(t)}=c_1(t)\ket{1}+c_2(t)\ket{2}$), and the occupation here is computed as (when $|V_0|\ll\hbar\omega_{12}$):
\begin{align}
    \mathcal{O}(t,\ket{2})=\left|c_2(t)\right|^2 = \frac{|V_0|^2}{\hbar^2}\frac{e^{-2\eta t}-2e^{-\eta t}\cos{(\omega_{12}-\omega)t}+1}{(\omega_{12}-\omega)^2 +\eta^2} \label{eq2}
\end{align}

\noindent When $t\rightarrow\infty$, we have:
\begin{align}
     \mathcal{O}(t=\infty,\ket{2})= \frac{|V_0|^2}{\hbar^2}\frac{1}{(\omega_{12}-\omega)^2 +\eta^2} \label{eq3}
\end{align}

To make the comparison easy, we will use equ.3, the final transition amplitude, instead of the transition rate, as the fundamental quantity to compare different methods. For the TLS calculation, we have used $\hbar\omega_{12}$ around 0.03 eV, $V_0$ around 0.0001 eV, and $\eta$ is taken as 0.001 fs$^{-1}$. These energy scales are chosen to be consistent to a typical phonon excited electronic transition problem. But as we will discuss later, scaling rule can be used to extend the conclusion to other values of these parameters. 
The final state $\ket{2}$ occupation ($\mathcal{O}(t=\infty,\ket{2})$) is taken at large $t$ to ensure the vanishment of the off-diagonal elements of the Hamiltonian. Here, the temperature is set to be very high to remove the detailed balance effect on state transition. This is in the same spirit of the original FGR, where there is no temperature dependent detailed balance effect.

\begin{figure}
\centering
\includegraphics[width=1\columnwidth]{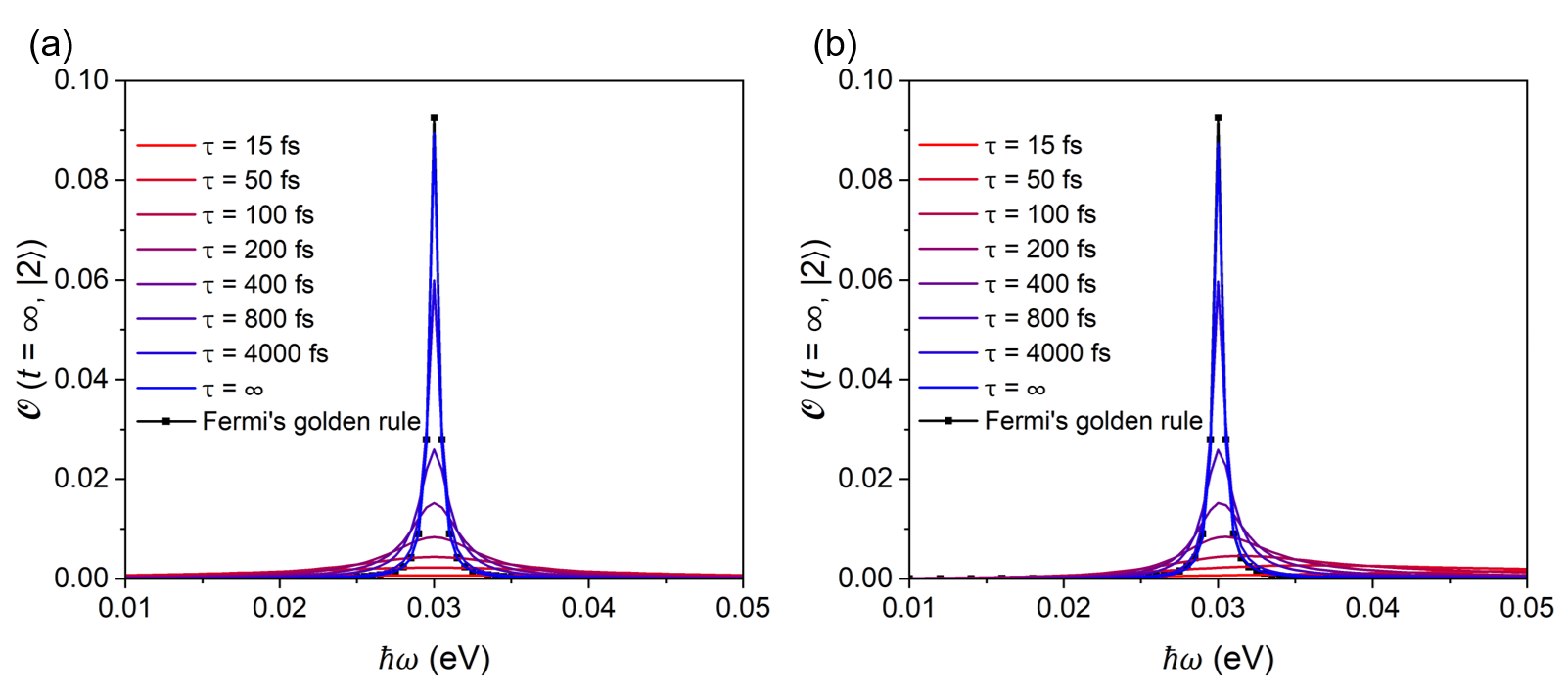}
\caption{State $\ket{2}$ occupation at $t=\infty$ for various frequencies ($\omega$) of the external perturbation but with fixed state energy different ($\hbar\omega_{12}=0.03$ eV). The initial state is a fully occupied state on $\ket{1}$. Comparisons of FGR and the numerical method calculation with (a) the fixed-basis and with (b) the adiabatic basis.} 
\label{fig1}
\end{figure}

\textit{Fixed basis.} Shown in Fig.1(a) is the calculated $\mathcal{O}(t=\infty,\ket{2})$ by using the numerical method with a fixed basis as a function of frequencies $\omega$ of external fields. The details of simulation are in SI 3.1. Various $\tau$ are computed for comparison and they are also compared to FGR by equ.3. When $\tau$ is infinite ($\tau=\infty$), calculated $\mathcal{O}(t=\infty,\ket{2})$ matches that of FGR. They show a symmetric distribution over $\hbar\omega=\hbar\omega_{12}=0.03$ eV. However, when a finite $\tau$ is applied in simulations, $\mathcal{O}(t=\infty,\ket{2})$ starts to deviate from FGR calculation. The distribution of $\mathcal{O}(t=\infty,\ket{2})$ is still symmetric around the resonant energy, but the maximum $\mathcal{O}(t=\infty,\ket{2})$ decreases with the decrease of the $\tau$ accompanied by spectrum line broadening. Moreover, we can integrate the occupation over the excitation frequency ($I(\tau)=\int \mathcal{O}(t=\infty,\ket{2}) d\hbar\omega$). We find that all the decoherence time including FGR yield the same total excitations ($I(\tau)=\int \mathcal{O}(t=\infty,\ket{2}) d\hbar\omega \approx 9.8\times10^{-5}$eV), which demonstrates the fulfillment of oscillator strength sum rule. In order to understand the results, we have derived the analytical expression of $\mathcal{O}(t=\infty,\ket{2})$. The detailed derivation is shown in SI S3.1, where the higher order term ($\mathcal{O}\left(V^4\right)$) is also derived. This is achieved by solving the partial differential equations (SI equ.3) with the boundary condition of the density matrix $D_{11}(t=0)=1,\ D_{22}(t=0)=0$, and the approximation of $\abs{V_0}\ll \hbar\omega_{12}$. The state $\ket{2}$ occupation (i.e. $D_{22}(t=\infty)$) is:
\begin{align}
    \mathcal{O}(t=\infty,\ket{2}) \equiv D_{22}(t=\infty) =\frac{\vert V_0\vert^2}{\hbar^2}\frac{1}{\left(\omega-\omega_{12}\right)^2+\left(\eta+\frac{1}{\tau}\right)^2}
\end{align}
The excitation by equ.4 agrees with the numerical method exactly (SI Fig.3). By comparing the FGR (equ.3), it is found that the decoherence effect will further broaden the FGR. The exponential decay of the coherence has the same consequence of decay of external field, where additional frequencies of stimulus is applied to the system. Such decoherence induced transitions can also be observed from the perspective of transition rate, as shown in SI S3.3.

\textit{Adiabatic basis.} Here, we implement the P-matrix NAMD simulation to evolve the density matrix and solve $D_{22}(t=\infty)$ (see calculation details in SI S1). When there is no decoherence, the P-matrix yields the same result as the analytical TDSE used to derive the FGR, as shown in SI Fig.1. The decoherence effect is introduced in the P-matrix method by a decaying term of the off-diagonal density matrix element on the basis of the adiabatic states. An additional term could also be introduced to incorporate the detailed balance~\cite{Kang19p224303,Zheng19p6174}, although it has no effect in the current study. The P-matrix result also agrees with other NAMD approach results as shown in SI Fig.2. Thus, the conclusion of this study does not depend on what NAMD approach is used, although the P-matrix method significantly accelerates our simulation speed and makes the current study more feasible.

Shown in Fig.1(b) is the simulated $\mathcal{O}(t=\infty,\ket{2})$ with the adiabatic basis. Similar to the fixed basis, it only matches FGR's results when the decoherence time is infinite. As the emerging of the decoherence effect (decreasing $\tau$), the maximum value of $\mathcal{O}(t=\infty,\ket{2})$ becomes suppressed and the distribution is broadened. In particular, when $\tau$ is very small, as shown in SI 3.3 from the perspective of transition rate $T_{ij}(t)$, the decoherence can introduce real carrier transitions in a broad range of $\omega$ before the external perturbation starts to play a role, such derivation is also similar in the case of fixed basis.
Meanwhile, the overall excitation ($I(\tau)=\int \mathcal{O}(t=\infty,\ket{2}) d\hbar\omega \approx 9.8\times10^{-5}$ eV) is also conserved in terms of $\tau$. However, different from the fixed basis, as the decoherence time decreases, the maximum $\mathcal{O}(t=\infty,\ket{2})$ frequency $\omega$ is shifted to higher energies, and the distribution of $\mathcal{O}(t=\infty,\ket{2})$ over $\hbar \omega=\hbar\omega_{12}$ becomes asymmetric. The peak frequency shift and the asymmetry originate from the nature of non-adiabatic coupling where the ratio of $\omega$ and $\omega_{12}$ will play a role in addition to the external field (see SI 3.2). A finite $\tau$ will further strengthen such shift and asymmetry.
The difference from FGR becomes more distinct in Fig.2. In this figure, the frequency of external perturbation is fixed as $\hbar\omega=0.03$ eV, but $\hbar\omega_{12}$ is varied. Our calculation shows that under a given external stimulation frequency $\omega$, if there are many different fixed states $\ket{2}$ with different energies $\epsilon_2$, for a finite $\tau$, it is not the one with the resonant frequency ($\hbar\omega=\hbar\omega_{12}$) that will have the maximum transition (or absorption). Instead, the $\hbar\omega_{12}$ below the resonant energy can have a larger absorption, particularly, when $\tau$ is small. In order to understand such behavior, we also try to solve $D_{22}(t=\infty)$ analytically. Here, based on the Hamiltonian in equ.1, the adiabatic states can be obtained exactly by diagonalizing the Hamiltonian matrix. Then, the non-adiabatic coupling $\left<\phi_{i}(t)\middle|\Dot{\phi}_{j}(t)\right>$ is solved as $1/2\hbar\omega\sin{\theta}e^{i\omega t}$ where $\sin{\theta}=\frac{4V_0}{\sqrt{4V_0^2+\left(\epsilon_1-\epsilon_2\right)^2}}$. By employing the approximation $|V_0|\ll\hbar\omega_{12}$, we have:
\begin{align}
      \mathcal{O}(t=\infty,\ket{2}) \equiv D_{22}(t=\infty) &= \left( \frac{\omega V_0}{\hbar\omega_{12}} \right)^{2} \frac{1+\frac{1}{\tau \eta}}{\left(\omega-\omega_{12}\right)^{2}+\left(\frac{1}{\tau}+\eta\right)^{2}}.
\end{align}
The detailed derivation is shown in SI S3.2, where the higher order contribution up to $\mathcal{O}(V^4)$ is also derived. This excitation by equ.5 matches with the P-matrix method exactly (SI Fig.4). In Fig.2, when $\hbar\omega_{12}$ is small, such large transition is mainly caused by the decoherence effect. This can be understand from the picture of wavefunction collapsing intuitively. When these eigenstates already mix the states $\ket{1}$ and $\ket{2}$, the wavefunction collapsing to the eigen states ($\ket{\phi_1}$ and $\ket{\phi_2}$) will enforce the carrier population to state $\ket{2}$, although such transition does not follow FGR. In the above sections, we have investigated two types of basis: fixed basis and adiabatic basis. In reality, choosing which basis is out of scope of this work. However, for most microscopic systems with quantum limit of decoherence, the eigen states (i.e. adiabatic states) are often the preferred basis\cite{PazJuanPablo1999QLoD}. 

\textit{Multi-external stimulations.} The derivation of single external stimulation $V_{0}e^{i\omega t}$ in equ.1 can be easily extended to multiple external stimulations: $\sum_{s}V_se^{i\omega_{s}t}$. This is more common in real materials where multiple phonons could contribute simultaneously to excite carriers between two eigen states. Furthermore, other stimulation, e.g. light, could perturb carrier excitation in addition to the phonons. In SI S3.4, we have derived the expression of $\mathcal{O}(t=\infty,\ket{2})$ up to $\mathcal{O}(V^2)$ under the multi-external stimulations with the adiabatic basis. Particularly, for TLS with two simultaneous phonons $V_1e^{i\omega_1 t} +V_2e^{i\omega_2 t} $, the final expression is simply the excitation by  $V_1e^{i\omega_1 t}$, the excitation by $V_2e^{i\omega_2 t}$, and their interference terms (SI equ.28).

\begin{figure}[H]
\centering
\includegraphics[width=1\columnwidth]{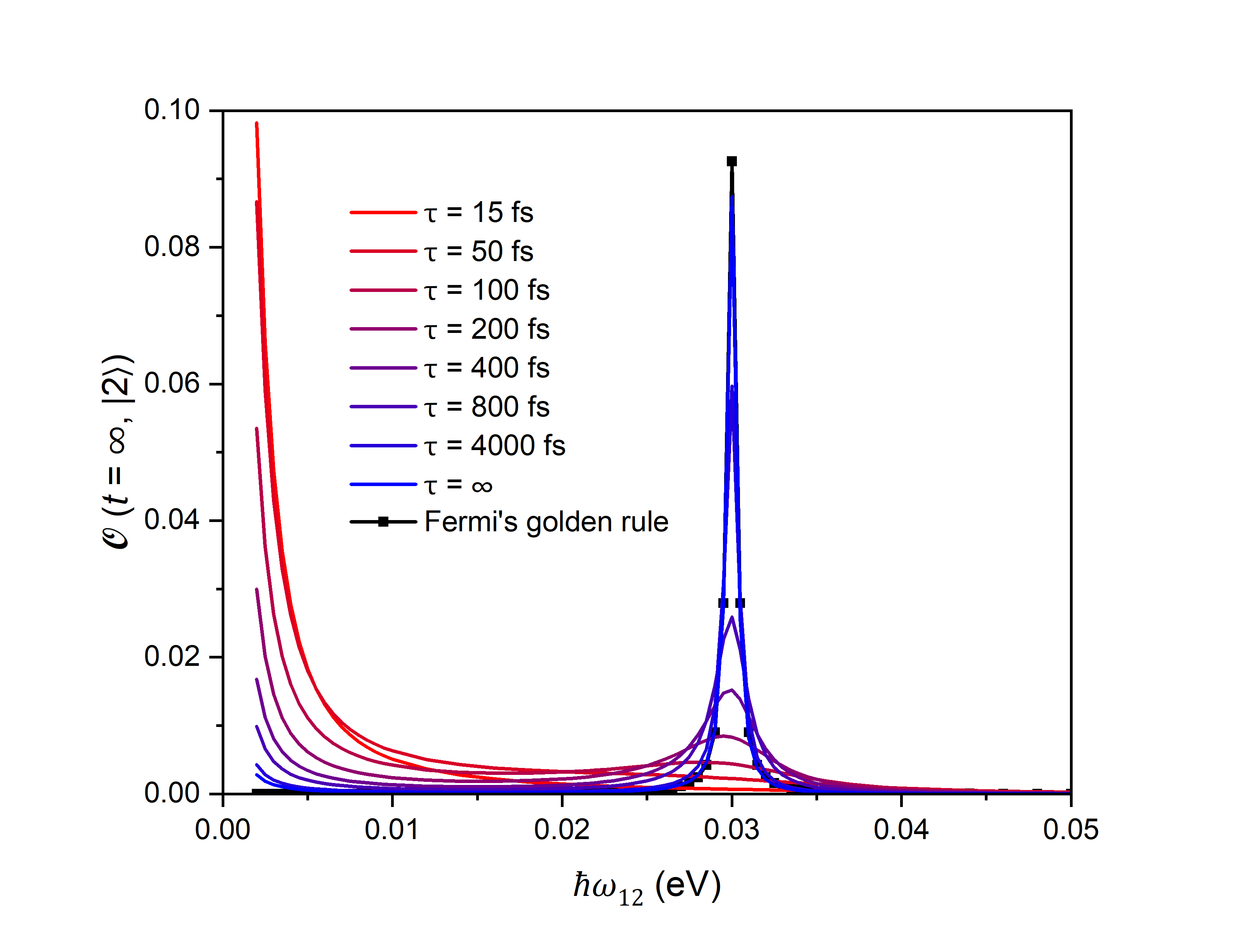}
\caption{State $\ket{2}$ occupation at $t=\infty$ for the various gap energies $\hbar\omega_{12}$ with a fixed frequency ($\hbar\omega=0.03$ eV) of the external perturbation. The initial state is a fully occupied state on $\ket{1}$. } 
\label{fig2}
\end{figure}

\textit{A real example: monolayer ${WS_2}$.} We investigate a real physical system $\mathrm{WS_2}$ to understand the role of decoherence to the carrier excitation with the first-principle method. For a $6\times6\times1$ supercell of monolayer $\mathrm{WS_2}$, we are interested in the carrier transitions between three states: VBM, VBM-1, and VBM-2 (SI S4.2). In this supercell, VBM-2 is from $\mathbf{k}=\Gamma$ point of the unit cell. VBM-1 and VBM are degenerate and they are from $\mathbf{k}=\pm\left(\frac{1}{3},\frac{1}{3},0\right)$ of the unit cell folded to the supercell band structure. The energy differences of these three states are in the similar range of the TLS aforementioned. Particularly, we are interested in the transitions of holes from VBM-2 to VBM (and VBM-1), which are modulated by the phonon modes with wavevector $\mathbf{q}=\pm\left(\frac{1}{3},\frac{1}{3},0\right)$ through the EPC. By implementing P-matrix formalism, it is possible to obtain the hole population at the VBM-1 and VBM states through the first-principle P-matrix NAMD simulation after placing a hole at VBM-2. Here, decoherence time $\tau$ is an input parameter. 
Furthermore, the energy difference between VBM-2 and VBM-1 can be tuned by applying hydrostatic stress in xy plane (VBM-1 and VBM remain degenerate independent of stress). 
Such energy gaps increase from 0.056 eV (tensiled structure) to 0.067 eV (stress-free structure), to 0.12 eV (compressed structure) (SI Fig.8).
This NAMD simulation is a post-processing of the ground-state first-principle molecular dynamics performed at 100 K for this system (see calculation details in SI S4.3). The details of first-principle P-matrix can also be found in Ref.20 and Ref.21. Shown in Fig.3 is the simulated excited holes at VBM-1 and VBM together as a function of $\tau$ with different energy gaps.
For a given energy gap, the decoherence illustrates a profound impact to the hole excitation compared to FGR. For small $\hbar\omega_{12}$, excited occupation has a relatively small variance unless $\tau$ is extremely small. However, for large $\hbar\omega_{12}$, the occupation has a sharp peak. In order to understand such behavior, we re-map the three-state system to the aforementioned TLS and perform NAMD simulations. The $\mathrm{WS_2}$ three-state Hamiltonian is simplified as SI equ.29. Since we are only interested in the transitions from VBM-2 to VBM-1 and VBM, a unitary transformation is applied to yield an effective TLS Hamiltonian (SI S5). Here, the coupling strength ($V$) is directly obtained using density functional perturbation theory (DFPT) method by computing the EPC matrix element at $\mathbf{q}=\pm\left(\frac{1}{3},\frac{1}{3},0\right)$ of the unit cell of $\mathrm{WS_2}$. The frequency of the external perturbation ($\omega$) is taking from the DFPT calculated phonon frequency (see calculation details in SI S6).

\begin{figure}
\centering
\includegraphics[width=1\columnwidth]{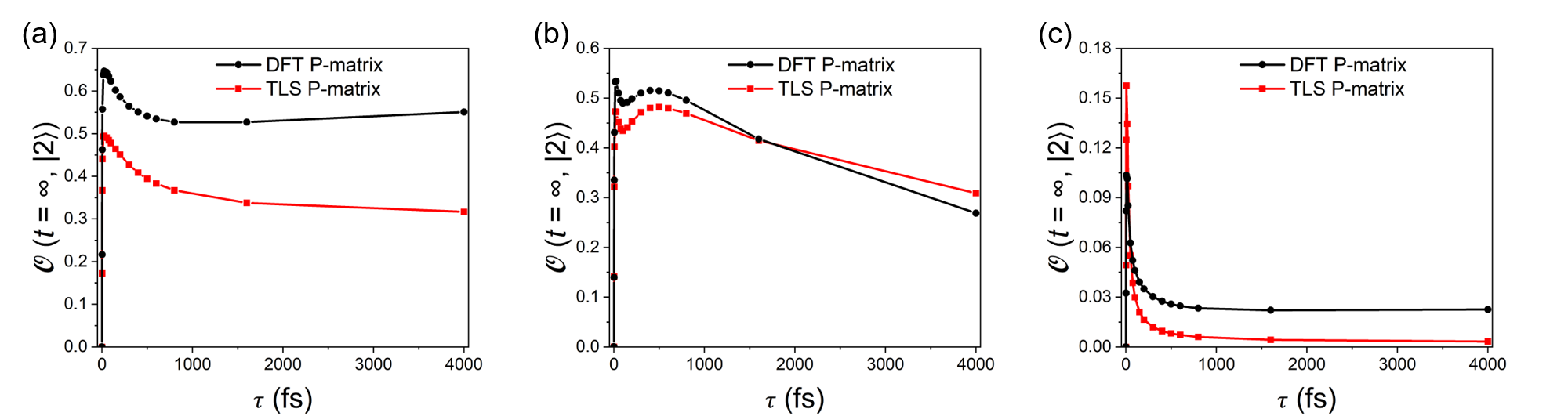}
\caption{The NAMD-simulated excited carrier of the real system and the TLS. Two phonons are considered, whose phonon energies are 0.046 and 0.020 eV. The corresponding EPC strength are 0.007 and 0.016 eV. (a) Tensiled structure ($1\%$ tensile): energy gap is 0.056 eV. (b) Stress-free structure: energy gap is 0.067 eV. (c) Compressed structure ($0.4\%$ compression): energy gap is 0.12 eV. The initial state is a fully occupied state on VBM-2.}
\label{fig3}
\end{figure}

Shown in Fig.3 is the P-matrix simulated $\mathcal{O}(t=\infty,\ket{2})$ based on the effective TLS to represent transitions from VBM-2 to VBM-1
(and VBM) in $\mathrm{WS_2}$. For three different energy gaps, we find that such TLS can well-reproduce the DFT trend. Here, for the tensiled structure, the effective TLS yields a smaller $\mathcal{O}(t=\infty,\ket{2})$ than that of DFT. This is because for a large coupling strength limit, the three-state system (DFT case) can pump $\frac{2}{3}$ electrons to upper state (i.e. $\frac{1}{3}$ electrons of each state when $t=\infty$) and the TLS can at most excite 0.5 electrons. It is also interesting to note that for the $\mathbf{q}$ phonon modes that couple these two states (monolayer $\mathrm{WS_2}$ unit cell has 3 atoms and 9 modes at this $\mathbf{q}$), two of them will show relatively large coupling strengths. However, taking any $\textit{one}$ out these 9 modes to fit will not match the DFT result at all (see SI Fig.9). At least the two modes with large coupling must be used together. Here, we want to comment that when the energy gap of two states is large compared to all the phonon modes (e.g. Fig.3(c)), the excited occupation has a strong dependent on the decoherence time (a sharp peak). This situation could be common in a molecular system where the state-energy difference is usually larger than the vibration energy. The decoherence time should be chosen carefully particularly near the sharp peak in real calculations. However, when the energy gap is small and close to the EPC strength, particularly when multi-phonon is contributing simultaneously (this is common in solid-state materials e.g. Fig.3(a)), the overall excited occupation will have a relatively week dependence on the decoherence time. This may explain partially why many calculations of periodic materials have revealed that carrier dynamics in a periodic material (e.g. hot carrier cooling,\cite{KangJun2019Nmdw,ZhengFan2023MkPN}carrier transitions,\cite{LiWei2021Ainm,ZhengCaihong2024CTi2, TangYue2025Sace} etc) do not change significantly by using a different decoherence time.

\begin{figure}
\centering
\includegraphics[width=1\columnwidth]{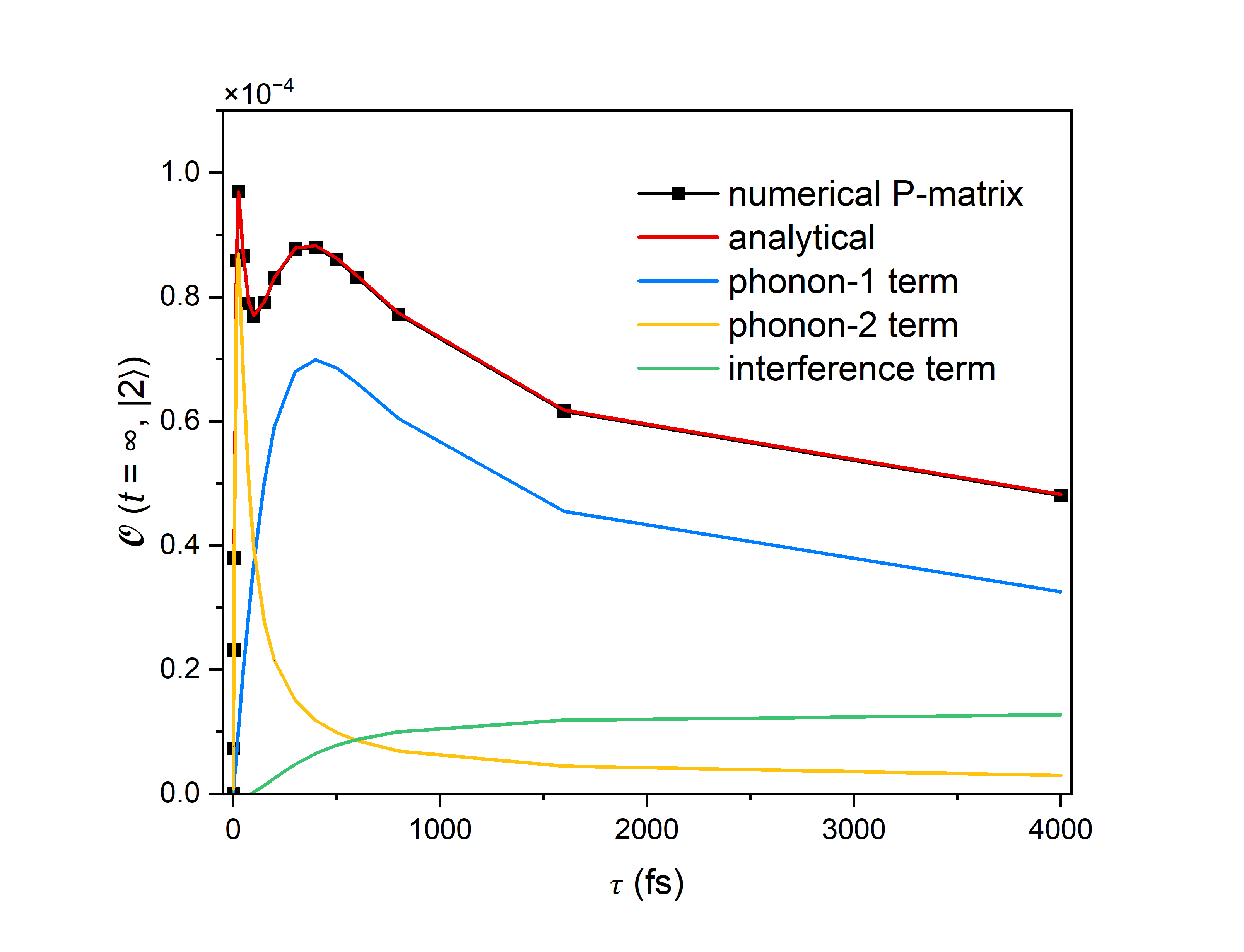}
\caption{State $\ket{2}$ occupation at $t=\infty$ of the two-phonon TLS model: phonon energies $\hbar \omega_{1}$ = 0.047 eV, $\hbar \omega_{2}$ = 0.020 eV; EPC strengths $V_1$ = 0.00001 eV, $V_2$ = 0.0001 eV; energy gap $\abs{\Delta}$ = 0.056 eV. Black: the numerical result of the P-matrix NAMD method.
Red: analytical result based on SI equ.28.
Blue: analytical result of the first phonon ($\hbar \omega_{1}$, $V_1$).
Gold: analytical result of the second phonon ($\hbar \omega_{2}$, $V_2$).
Green: analytical result of the interference terms.}
\label{fig4}
\end{figure}
Due to the large coupling strength, it is unable to understand with the analytical expression where an approximation of $|V|\ll \hbar\omega_{12}$ must be present. Here, by using the stress-free structure as an example where the curve shows a two-peak feature (Fig.3(b)), we scale down the coupling strengths of these two phonons and recalculate the TLS model. Shown in Fig. 4 are the numerical P-matrix and analytical results, which match with each other exactly. Based on the analytical expression (SI equ.28), the transition is dominated by one phonon (with un-resonant frequency but a large EPC strength) when the decoherence time is short, while for long decoherence time, the transition is dominated by the other phonon (whose energy is close to the gap).

In summary, we use a TLS as an example to compare the carrier excitations calculated by NAMD and FGR. We find that only when the decoherence time used in NAMD is infinite, these two methods give the same results. We also investigate two types of basis: the fixed basis and the adiabatic basis. The impact of decoherence effect is different for these two basis. However, they both deviate from the FGR significantly when the decoherence time becomes short. We further demonstrate such decoherence effect in a real system of monolayer $\mathrm{WS_2}$ with the first-principle calculations and they gives consistent results. Our result can provide guidance to state transition problems from quantum computing to ultra fast carrier dynamics when the decoherence effect is significant.
    
\noindent \textbf{\large{Acknowledgment}}

This work is supported by the National Natural Science Foundation of China (62305215) and ShanghaiTech AI4S initiative SHTAI4S202404. The computational support is provided by the HPC platform of ShanghaiTech University. We want to thank Dr. Lin-Wang Wang for his insightful contribution and advice for this manuscript and the project as a whole.

\bibliography{manuscript}

\end{document}


\captionsetup[figure]{name={SI Fig.}}

\maketitle

\textbf{S1. Computation details of P-matrix NAMD and comparison between P-matrix NAMD and exact solution for solvable two-level model system (TLS).} 

In P-matrix formalism, for every ``MD step", the adiabatic inner product is computed as $U_{ij}(t,t+\Delta t)=\left<\phi_i(t)|\phi_j(t+\Delta t)\right>$, where $\Delta t$ is the MD step length ($\Delta t=$ 2 fs), $\phi_{i}(t)$ is the adiabatic states diagonalized at MD step. By using $U_{ij}$ and time-dependent eigen energies as inputs, the Hamiltonian is re-built and interpolated linearly within one MD step. A set of diagonalizations are then performed based on the interpolated Hamiltonian. Between the two diagonalizations, the density matrix is evolved following Liouville–von Neumann equation with small time-step $dt$ ($dt$ around 0.1 a.s.). Here, 10 diagonalizations within one MD step and 1000 small $dt$ between two diagonalizations are used to converge the result. 

Based on the Hamiltonian in equ.1 in the main text, such TLS system can be solved exactly to obtain time-dependent occupations of state $\ket{1}$ and $\ket{2}$~\cite{Sakurai17p}:
\begin{align}
    |c_2(t)|^2 & = \frac{V_0^2/\hbar^2}{V_0^2/\hbar^2+(\omega-\omega_{12})^2/4}\sin^2
    \left\{\left[\frac{V_0^2}{\hbar^2}+\frac{(\omega-\omega_{12})^2}{4}\right]^{1/2}t\right\} \nonumber \\
    |c_1(t)|^2 & = 1-|c_2(t)|^2
\end{align}

Shown in SI Fig.1 is the comparison between P-matrix (no decoherence and very high temperature) calculated state $\ket{2}$ occupation and the exact solution. These two methods give exactly the same results.
\begin{figure}[H]
\centering
\includegraphics[width=0.8\columnwidth]{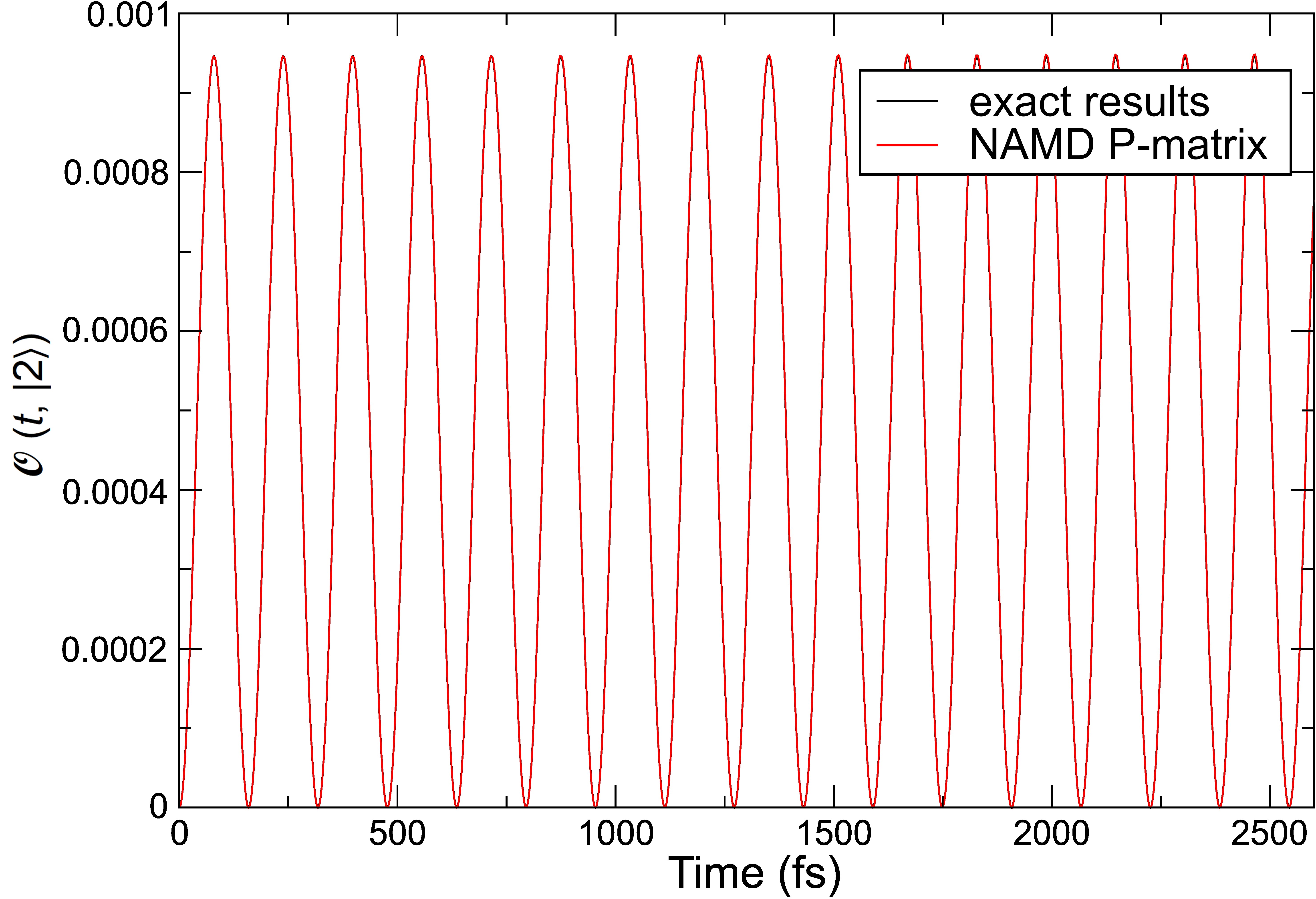}
\caption{State $\ket{2}$ occupation as a function of time computed by the NAMD P-matrix and the exact solution. These two methods give the same results. }
\label{sfig1}
\end{figure}

\textbf{S2. Comparison between P-matrix and DISH.} 

Shown in SI Fig.2 is the comparison between P-matrix and DISH calculated state $\ket{2}$ occupation at $t=\infty$ ($\mathcal{O}(t=\infty,\ket{2})$) for different state energy differences but with a fixed external perturbation frequency $\hbar\omega=0.03$ eV. The decoherence time $\tau=50$ fs is tested. Here, decoherence scheme used in DISH is implemented based on Ref.~\citenum{Jaeger12p22A545}. The decoherence event occurs based on a Poisson distribution with average decoherence time as 50 fs. The non-adiabatic dynamics is simulated similar to P-matrix but a wavefunction instead of density matrix is used for evolution. 10 diagonalizations are performed within one MD step and 1000 small wavefunction evolution steps are used between two diagonalizations. The final DISH result is averaged over 6000 stochastic trajectories for each calculation. As shown in SI Fig.2, P-matrix and DISH show general agreement, especially for small $\hbar\omega_{12}$ where state $\ket{2}$ occupation is large. More trajectories maybe needed to reach better agreement for the low state-$\ket{2}$ occupations.
\begin{figure}[H]
\centering
\includegraphics[width=0.8\columnwidth]{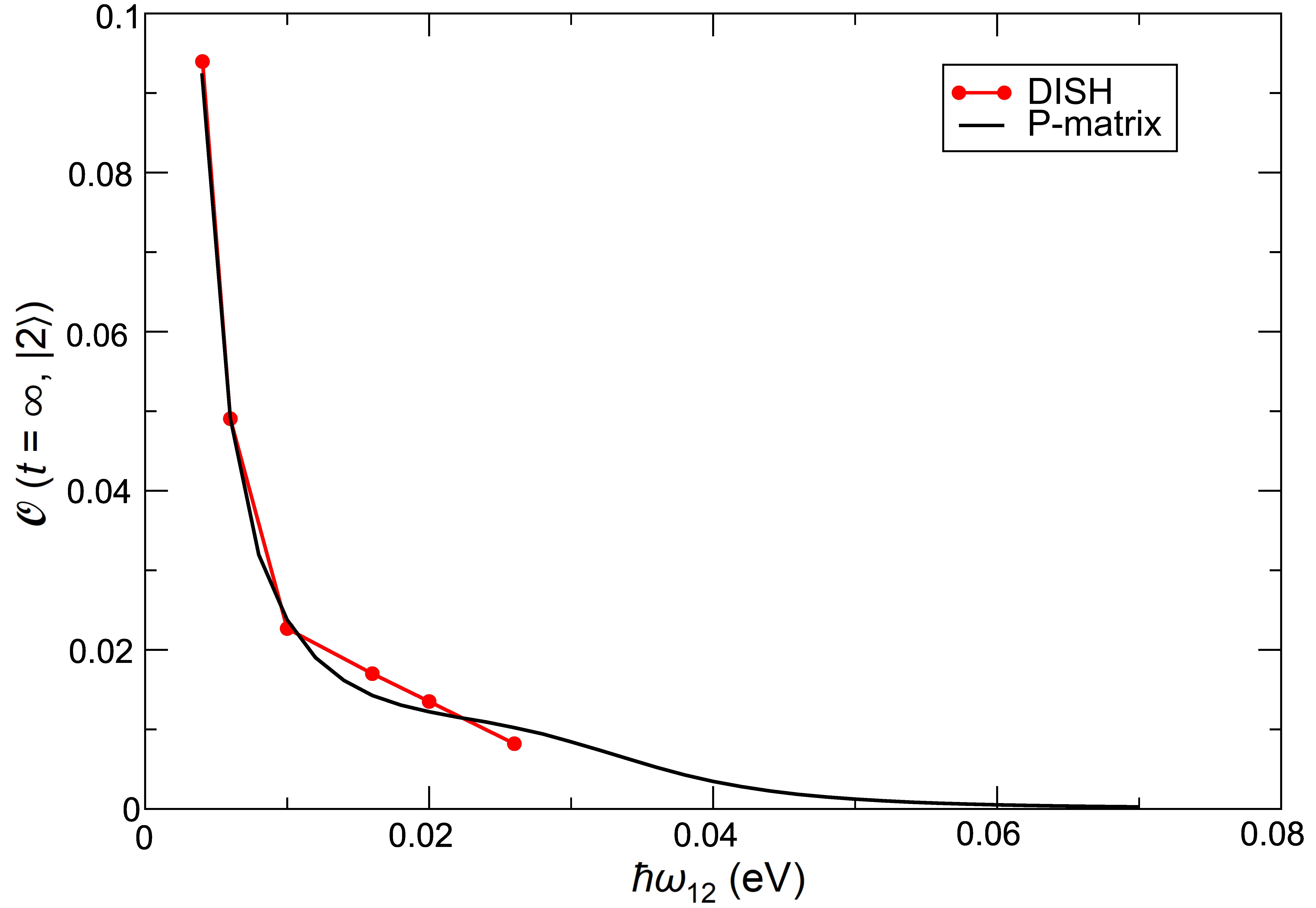}
\caption{DISH and P-matrix methods comparison for state $\ket{2}$ occupation at $t=\infty$ ($\mathcal{O}(t=\infty,\ket{2})$) as a function of state energy difference $\hbar\omega_{12}$ with a fixed external perturbation frequency ($\hbar\omega=0.03$ eV). For DISH, calculations with $\hbar\omega_{12}=$ 0.004, 0.006, 0.010, 0.016, 0.020, 0.026 eV are performed. The initial state is a fully occupied state on $\ket{1}$. Decoherence time $\tau=50$ fs. }
\label{sfig2}
\end{figure}

\textbf{S3. Derivation for decoherence effect to Fermi's golden rule.}

\textbf{S3.1: Fixed basis.}

By using the fixed states $\ket{1}$ and $\ket{2}$ as the basis throughout the whole time evolution, the density matrix $D(t)$ follows the Liouville equation as (atomic unit is applied):
\begin{align}
    \mathbf{\dt{D}}(t) & = -\frac{i}{\hbar} \left[\mathbf{H}(t), \mathbf{D}(t)\right] 
\end{align}

\noindent where $\mathbf{H}(t)$ is the time-dependent Hamiltonian of the TLS (equ.1 in the main text). When the decoherence effect is included, the time evolution of density matrix $D(t)$ satisfies the following equations,
\begin{align}
    \dt{D}_{11}(t) & = -\frac{i}{\hbar} (H_{12}(t)D_{21}(t) - D_{12}(t)H_{21}(t)) \nonumber\\
    \dt{D}_{22}(t) & = -\frac{i}{\hbar} (H_{21}(t)D_{12}(t) - D_{21}(t)H_{12}(t))  \nonumber \\
    \dt{D}_{12}(t) & = -\frac{i}{\hbar} (H_{11}(t)D_{12}(t) - D_{11}(t)H_{12}(t) + H_{12}(t)D_{22}(t) - D_{12}(t)H_{22}(t))  - \frac{D_{12}(t)}{\tau}  \label{eq3}
\end{align}
\noindent and $D_{21}(t)=D^*_{12}(t)$. All the elements of the above density matrix and the Hamiltonian are time-dependent. The occupation on state $\ket{2}$ after a long time will be $D_{22}(t=\infty)$.

We evolve equ.3 through long-time numerical integration to obtain the numerical result of $D_{22}(t=\infty)$ and the result is shown in the main-text Fig1.(a). Furthermore, we employ an analytical approach to derive the analytical result of $D_{22}(t=\infty)$. By plugging the elements of the Hamiltonian, we obtain the coherence term in equ.3 as:
\begin{align}
    \dt{D}_{12} = -\frac{i}{\hbar}V_0e^{i\omega t -\eta t}\left(D_{22}-D_{11}\right) + \frac{i}{\hbar}\omega_{12}D_{12} - \frac{D_{12}}{\tau}.
\end{align}

\noindent This is a partial differential equation in terms of time $t$, which yields:
\begin{align}
    D_{12}(t) = \frac{V_0/\hbar}{\omega-\omega_{12}+i(\eta-1/\tau)} \left(e^{i\omega_{12}t-t/\tau}-e^{i\omega t-\eta t}\right) \left(D_{22}-D_{11}\right).
\end{align}

\noindent Now, by plugging the above expression of $D_{12}(t)$ and $D_{11}+D_{22}=1$ for any time $t$ into the evolution of $D_{22}$ in equ.3, we have
\begin{align}
    \dt{D}_{22} &= 2\frac{V_0}{\hbar^2} e^{-\eta t} \underbrace{\operatorname{Im}\left[\frac{V_0}{\omega-\omega_{12}+i(\eta-1/\tau)}\left(e^{-i\left(\omega-\omega_{12}\right)t-t/\tau}-e^{-\eta t}\right)\right]}_{I(t)}\left(2D_{22}-1\right) ,\nonumber\\
    &\equiv 4\frac{V_0}{\hbar^2} e^{-\eta t} I(t) D_{22} - 2\frac{V_0}{\hbar^2}e^{-\eta t} I(t).
\end{align}

\noindent This is a more complex partial differential equation, and it can be solved with the approximation of small $V_0$. Since $D_{22}$ is small and $I(t)$ is on the order of $V_0$, if only consider the expansion of $D_{22}(t)$ up to $V_0^2$-order, a much simpler differential equation can be obtained by ignoring the first term on the right-hand side of the above equation. In this case, we can solve that 
\begin{align}
    D_{22}(t) \approx & -\frac{V_0^2/\hbar^2}{\left[\left(\omega-\omega_{12}\right)-i(1/\tau+\eta)\right]\left[(\omega-\omega_{12})-i(1/\tau-\eta)\right] } e^{-i(\omega-\omega_{12})t-(1/\tau+\eta)t} \nonumber\\
              & - \frac{V_0^2/\hbar^2}{\left[\left(\omega-\omega_{12}\right)-i(1/\tau+\eta)\right]\left[(\omega-\omega_{12})-i(1/\tau-\eta)\right] }
    e^{i(\omega-\omega_{12})t-(1/\tau+\eta)t} \nonumber\\
              & -\frac{V_0^2/\hbar^2\left(1/\left(\tau\eta\right) -1\right)}{(\omega-\omega_{12})^2+(1/\tau-\eta)^2} e^{-2\eta t} + C,
\end{align}

\noindent where $C$ is a constant which can be determined by the initial condition $D_{22}(t=0)=0$. For the long time limit ($t\rightarrow \infty$) with $e^{-\eta t}\rightarrow 0$, $D_{22}(t=\infty)=C$, which is
\begin{align}
    D_{22}(t=\infty) \approx \left(\frac{V_0}{\hbar}\right)^2\frac{ \frac{1}{\tau\eta} +1}{(\omega-\omega_{12})^2+(1/\tau+\eta)^2}.
\end{align}

Comparisons of numerical and analytical results based on fixed basis are shown in SI Fig.3. It can be seen that the analytical solution is consistent with the numerical solution.

Furthermore, a higher order in terms of $V_0$ could be obtained. By conserving the $D_{22}$ term in the right-hand side of the equ.6 and expanding up to the $V_0^4$-order, we obtain
\begin{align}
    D_{22}(t=\infty) \approx&\left(\frac{V_0}{\hbar}\right)^2 \frac{ \frac{1}{\tau\eta} +1}{(\omega-\omega_{12})^2+(1/\tau+\eta)^2} \nonumber\\
    &- \left(\frac{V_0}{\hbar}\right)^4\frac{1}{\left[(\omega-\omega_{12})^2+(1/\tau-\eta)^2\right]^2} \left[ \left(\frac{1}{\tau\eta}-1\right)^2 \right.\nonumber\\
     & \left. +\frac{8(1/\tau-\eta)\left[(\omega-\omega_{12})^2(3/\tau+\eta)-(1/\tau-\eta)(1/\tau+\eta)^2\right]}{\left[(\omega-\omega_{12})^2+(1/\tau+\eta)^2\right]\left[(\omega-\omega_{12})^2+(1/\tau+3\eta)^2\right]}\right]
\end{align}

\begin{figure}[H]
\centering
\includegraphics[width=0.9\columnwidth]{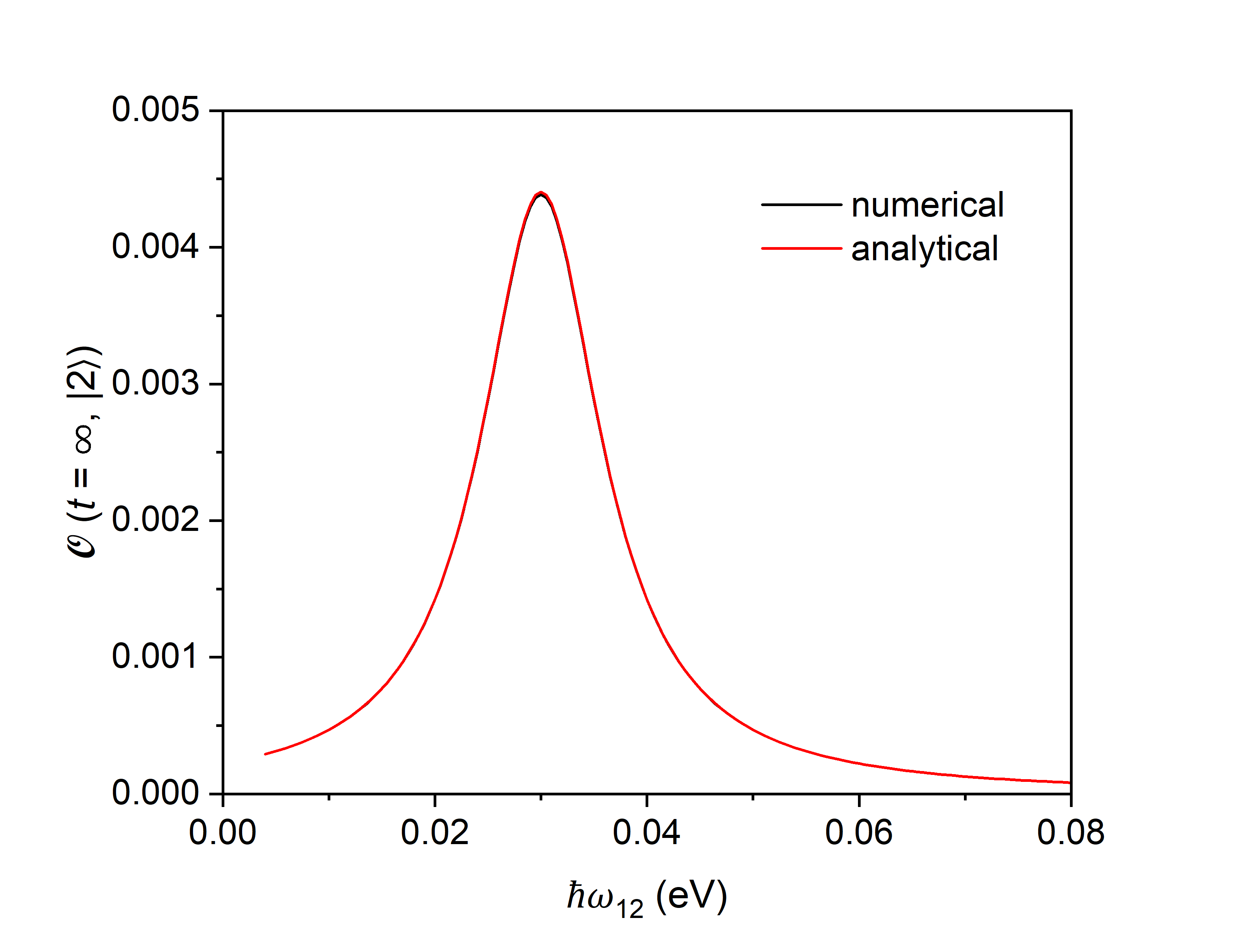}
\caption{Comparisons of numerical and analytical results (in terms of $V_0^2$-order) of state $\ket{2}$ occupation based on fixed basis at $t=\infty$ ($\mathcal{O}(t=\infty,\ket{2})$) as a function of state energy difference $\hbar\omega_{12}$ with a fixed external perturbation frequency ($\hbar\omega=0.03$ eV). The initial state is a fully occupied state on $\ket{1}$. Decoherence time $\tau=100$ fs and $V_0= 0.0001$ eV. }
\label{sfig3}
\end{figure}

For numerical solution, by expanding $\mathbf{H}(t)$ and $\mathbf{D}(t)$ as a $4\times4$ matrix and a 4-element vector, we have:
\begin{align}
    \frac{\partial}{\partial t}
    \left(
    \begin{array}{c}
         D_{11} \\
         D_{12}\\
         D_{21}\\
         D_{22}
    \end{array}
    \right)
    = \left(
    \begin{array}{cccc}
         0& iVe^{-i\omega t-\eta t} & -iVe^{i\omega t-\eta t}&0 \\
        iVe^{i\omega t-\eta t} &-i\left(\epsilon_1-\epsilon_2\right)-\frac{1}{\tau}&0&-iVe^{i\omega t-\eta t} \\
        -iVe^{-i\omega t-\eta t} &0& i\left(\epsilon_1-\epsilon_2\right)-\frac{1}{\tau}&iVe^{-i\omega t-\eta t} \\
         0 &-iVe^{-i\omega t-\eta t}&iVe^{i\omega t-\eta t}&0
    \end{array}
    \right)
    \left(
    \begin{array}{c}
         D_{11} \\
         D_{12}\\
         D_{21}\\
         D_{22}
    \end{array}
    \right)
\end{align}
which can be abbreviated as $\frac{\partial}{\partial t}D(t)=A(t)D(t)$. The Ruge-Kutta method is implemented to integrate this differential equation from the initial state $D(t=0)=[1,0,0,0]$. The total evolution time of $D(t)$ is about 16500 fs with the time step of 0.1 fs.

\textbf{S3.2: Adiabatic basis.}

When using the adiabatic states $\ket{\phi_1}$ and $\ket{\phi_2}$ as the basis, the time evolution of $D_{ij} (t)$ with the decoherence effect is included by satisfying the following equation,
\begin{align}
    \Dot{D}_{ij}(t) &= -\frac{i}{\hbar}\left[V(t),D(t) \right]_{ij} - \frac{D_{ij}(t)}{\tau_{ij}}.\ i,j=1,2;\ i \neq j.
\end{align}
where $V_{ij}$ is defined as $V_{ij}(t)\equiv\delta_{ij}\epsilon_{i}(t) -i\hbar\left<\phi_i(t)\middle|\dot{\phi}_j(t)\right>$ which is the non-adiabatic coupling matrix. The diagonal density matrix element is evolved as:
\begin{align}
    \Dot{D}_{ii}(t)&=-\frac{i}{\hbar}\left[V(t),D(t)\right]_{ii}
\end{align}

In this work, we implement the P-matrix NAMD simulation to evolve equ.11 and equ.12 and solve the $D_{22}(t=\infty)$ numerically. The P-matrix result is shown in the main-text Fig.1(b). Similarly, we derive the analytical expression of $D_{22}$. First, the non-adiabatic coupling matrix $V_{ij}$ needs to be expressed explicitly. For the Hamiltonian shown in the main-text equ.1, we have computed the eigenstates as:
\begin{align}
    \phi_1(t)=\begin{pmatrix}
\cos\frac{\theta}{2}\\
-\sin\frac{\theta}{2}e^{-i\omega t}
\end{pmatrix}, \phi_2(t)=\begin{pmatrix}
\sin\frac{\theta}{2}e^{i\omega t}\\
\cos\frac{\theta}{2}
\end{pmatrix},
\end{align}
where $\sin{\theta}=\frac{\abs{Ve^{-\eta t}}}{\sqrt{\left(Ve^{-\eta t}\right)^2+\Delta^2}},\ \cos{\theta}=\frac{-\Delta}{\sqrt{\left(Ve^{-\eta t}\right)^2+\Delta^2}}$, and $2\Delta=\epsilon_1-\epsilon_2$. Here, we assume $\epsilon_1<\epsilon_2$ and $\hbar\omega_{12}\equiv\abs{\epsilon_1-\epsilon_2}$.
With this solution, the non-adiabatic coupling matrix element $V_{12}(t)$ can be written as:
\begin{align}
    V_{12}(t)=\frac{\hbar\omega}{2}\sin{\theta}e^{i\omega t}
\end{align}
and the $\sin{\theta}$ can be approximated as:
\begin{align}
    \sin{\theta} &= \frac{2V_0}{\hbar\omega_{12}} e^{-\eta t}\frac{1}{\sqrt{1+\left(\frac{2V_0}{\hbar\omega_{12}} e^{-\eta t}\right)^{2}}} \approx \frac{2V_0}{\hbar\omega_{12}}e^{-\eta t} \left\{
    1-\frac{1}{2}\left(\frac{2V_0}{\hbar\omega_{12}} e^{-\eta t} \right)^{2}
    \right\}
\end{align}
with the approximation of $\abs{V_0} \ll\hbar\omega_{12} $.
In addition, we have:
\begin{align}
        V_{11}-V_{22}&=\epsilon_{1}(t)-\epsilon_{2}(t)
    \approx -\hbar\omega_{12} \left\{
    1+\frac{1}{2}\left(
    \frac{2V_0}{\hbar\omega_{12}}e^{-\eta t}
    \right)^{2}
    \right\}
\end{align}
By plugging the equ.15 and equ.16 into the evolution of $\dot{D}_{ij}$ with the first-order $V_0$ term, we have:
\begin{align}
    \Dot{D}_{12} + \left( -i\omega_{12} + \frac{1}{\tau} \right) D_{12}
    &= -i\frac{\omega V_0}{\hbar\omega_{12}}e^{-\eta t}e^{i\omega t} \Delta D,
\end{align}
where $\Delta D\equiv D_{22}-D_{11}$. This is also a partial differential equation in terms of time t, which yields:
\begin{align}
    D_{12}(t)
    &= \frac{\omega V_0}{ \hbar\omega_{12}} \frac{(\omega-\omega_{12})+i(\frac{1}{\tau}-\eta)}{\left(\omega-\omega_{12}\right)^{2} + (\frac{1}{\tau}-\eta)^{2}}e^{i\omega_{12} t-\frac{t}{\tau}} \Delta D - \frac{\omega V_0}{ \hbar\omega_{12}}\frac{\left(\omega-\omega_{12}\right)+i(\frac{1}{\tau}-\eta)}{\left(\omega-\omega_{12}\right)^{2} + (\frac{1}{\tau}-\eta)^{2}}e^{i\omega t - \eta t}\Delta D.
\end{align}
\noindent Now, by plugging the above expression of $D_{12}(t)$, $D_{11}+D_{22}=1$ and $
    V_{21} = V_{12}^{*}  \approx \frac{\omega V_0}{\omega_{12}} e^{-i\omega t-\eta t}
$
into the evolution of $D_{22}$ in the equ.12, we have:
\begin{align}
    \Dot{D}_{22} &= \frac{2}{\hbar}\rm{Im}\left(V_{21} D_{12}\right)\nonumber\\ 
   &\approx -2\left( \frac{\omega V_0}{\hbar\omega_{12}} \right)^{2}
    \frac{1}{\left(\omega-\omega_{12}\right)^{2} + \left(\frac{1}{\tau}-\eta\right)^{2}} \times \nonumber\\
     &\rm{Im}\left\{
    \left[ \left(\omega-\omega_{12}\right) + i\left(\frac{1}{\tau}-\eta\right) \right] e^{-i\left(\omega-\omega_{12}\right)t- \left(\frac{1}{\tau}-\eta\right) t} - \left[ \left(\omega-\omega_{12}\right) + i\left(\frac{1}{\tau}-\eta\right)  \right] e^{-2\eta t}
    \right\},
\end{align}
which can be solved on the order of $V_0^2$ as:
\begin{align}
     D_{22}(t= \infty) &\approx \left( \frac{\omega V_0}{\hbar\omega_{12}} \right)^{2} \frac{\frac{1}{\tau \eta} +1}{\left(\omega-\omega_{12}\right)^{2}+\left(\frac{1}{\tau}+\eta\right)^{2}}.
\end{align}

Furthermore, a higher order in terms of $V_0^4$ could be obtained by including the higher-order $V_0$ terms in equ.15 and equ.16:
\begin{align}
    D_{22}(t=\infty) &= \left(\frac{\omega V_0}{\hbar\omega_{12}}\right)^{2}\frac{1+\frac{1}{\tau \eta}}{\left(\omega-\omega_{12}\right)^{2}+\left(\frac{1}{\tau}+\eta\right)^{2}}\nonumber\\
    &- \omega^{2}\left(\frac{V_0}{\hbar\omega_{12}}\right)^4\left\{
    \frac{1+\frac{1}{\tau \eta}}{\left(\omega-\omega_{12}\right)^{2}+\left(\frac{1}{\tau}+\eta\right)^{2}}+\frac{3+\frac{1}{\tau \eta}}{\left(\omega-\omega_{12}\right)^{2}+\left(\frac{1}{\tau}+3\eta\right)^{2}}
    \right\}\nonumber\\
    &-8\left(\frac{\omega V_0}{\hbar\omega_{12}} \right)^{4} \left(\frac{1}{\left(\omega-\omega_{12}\right)^{2}+\left(\frac{1}{\tau}-\eta\right)^{2}}\right)^{2}\times\left\{
    \frac{\left(\frac{1}{\tau}-\eta\right)^{2}}{8\eta^{2}}\right.\nonumber\\
    &+\left. \frac{\left(\frac{1}{\tau}-\eta\right)\left[\left(\omega-\omega_{12}\right)^{4}-\left(\frac{1}{\tau}-\eta\right)\left(\frac{1}{\tau}+3\eta\right)^{2}\left(\frac{1}{\tau}+\eta\right) \right]}{2\eta\left[\left(\frac{1}{\tau}+\eta\right)^{2} + \left(\omega-\omega_{12}\right)^{2} \right]\left[\left(\frac{1}{\tau}+3\eta\right)^{2} + \left(\omega-\omega_{12}\right)^{2} \right]}\right.\nonumber\\
    &+\left.
    \frac{\left(5\eta+3\frac{1}{\tau} \right)\left(\frac{1}{\tau}-\eta \right)\left( \omega-\omega_{12}\right)^{2}}{\left[\left(\frac{1}{\tau}+\eta\right)^{2} + \left(\omega-\omega_{12}\right)^{2} \right]\left[\left(\frac{1}{\tau}+3\eta\right)^{2} + \left(\omega-\omega_{12}\right)^{2} \right]}\right.\nonumber\\
    &+\left.
    \frac{\left[\left(\omega-\omega_{12}\right)^{2}-\left(\frac{1}{\tau}-\eta \right) \left( \frac{1}{\tau}+\eta\right)\right]^{2}}{2\left[\left(\frac{1}{\tau}+\eta\right)^2+\left(\omega-\omega_{12}\right)^2  \right]^2}
    \right\}.
\end{align}

\begin{figure}[H]
\centering
\includegraphics[width=0.9\columnwidth]{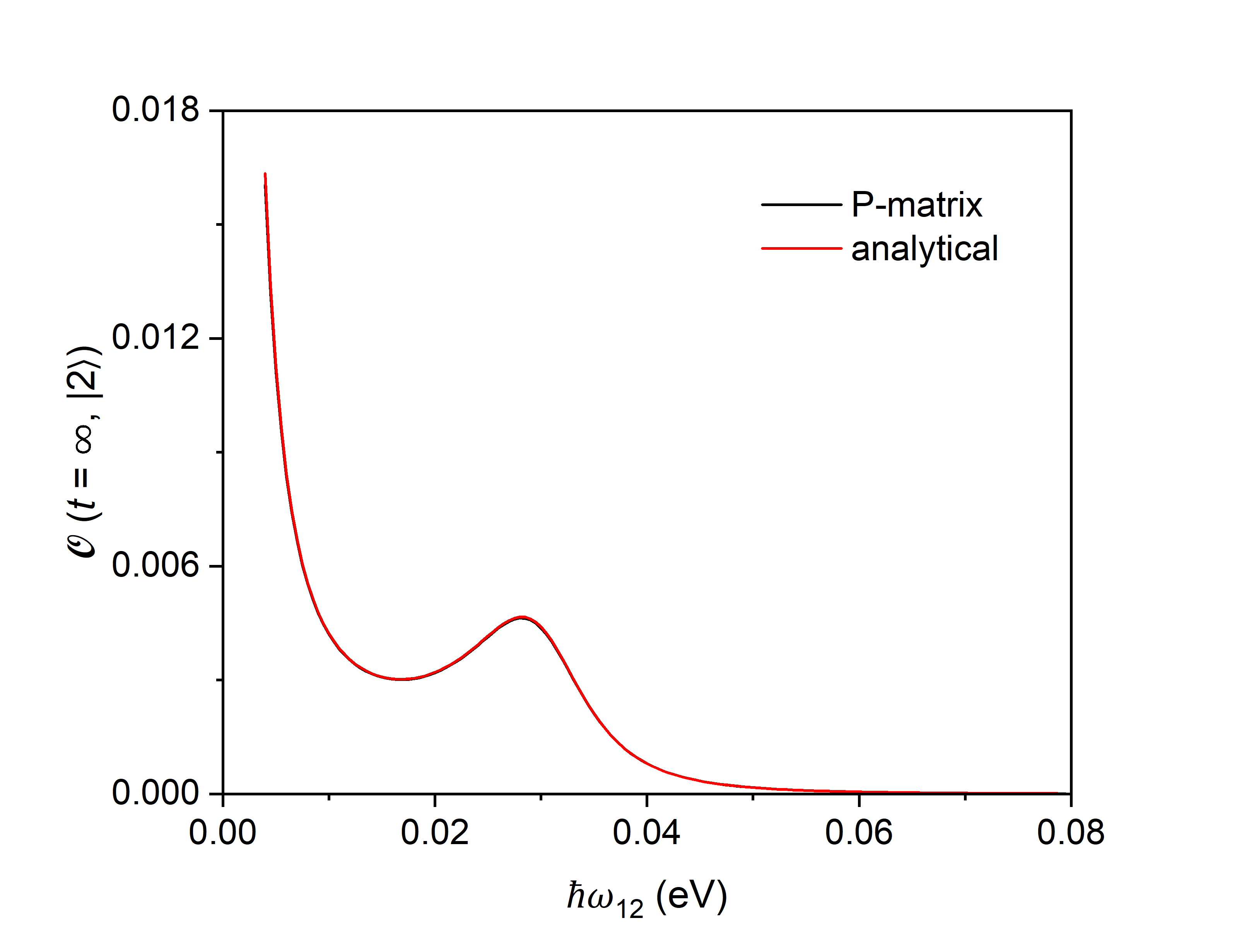}
\caption{Comparisons of numerical and analytical results (in terms of $V_0^2$-order) of state $\ket{2}$ occupation based on adiabatic basis at $t=\infty$ ($\mathcal{O}(t=\infty,\ket{2})$) as a function of state energy differences $\hbar\omega_{12}$ with a fixed external perturbation frequency ($\hbar\omega=0.03$ eV). The initial state is a fully occupied state on $\ket{1}$. Decoherence time $\tau=100$ fs and $V_0= 0.0001$ eV. }
\label{sfig4}
\end{figure}
Comparisons of numerical and analytical results based on adiabatic basis are shown in SI Fig.4 and SI Fig.5. It can be seen that when $V_0$ is small enough, the analytical solutions (both of the order in terms of $V_0^2$ and in terms of $V_0^4$) are consistent with the numerical solution. While $V_0$ is becoming large, the analytical solution of the order in terms of $V_{0}^{4}$ is still close to the numerical solution. But the analytical solution of the order in terms of $V_{0}^{2}$ yields a slight deviation. Here, we want to comment that the above derived $D_{22}(t=\infty)$ is still in the basis of adiabatic states, which has not been rotated to project to state $\ket{2}$ and compute its occupation as defined in the main text. But we know $V_0\ll \hbar\omega_{12}$, thus, the difference of projections to $\ket{2}$ and $\ket{\phi_2}$ is very small. Moreover, when $t\rightarrow\infty$, the external field is approaching zero. Such difference of projection is negligible (but not zero). Therefore, the derived $D_{22}$ can still match FGR derived with the fixed basis (equ.3) when $\tau\rightarrow \infty$.

\begin{figure}[H]
\centering
\includegraphics[width=0.9\columnwidth]{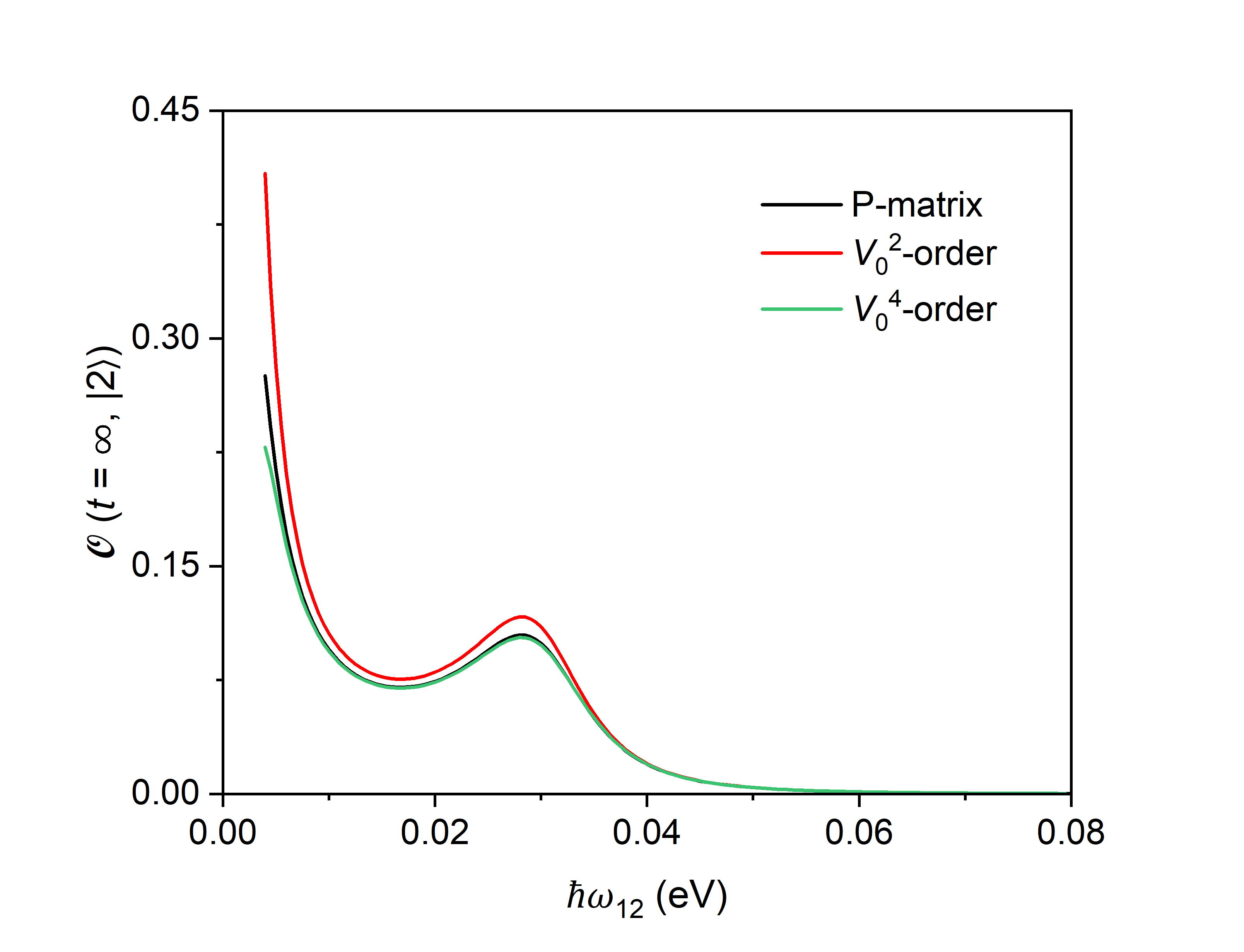}
\caption{Comparisons of numerical and analytical results (of the order in terms of $V_0^2$-order and in terms of $V_0^4$-order, respectively) of state $\ket{2}$ occupation based on adiabatic basis at $t=\infty$ ($\mathcal{O}(t=\infty,\ket{2})$) as a function of state energy differences $\hbar\omega_{12}$ with a fixed external perturbation frequency ($\hbar\omega=0.03$ eV). The initial state is a fully occupied state on $\ket{1}$. Decoherence time $\tau=100$ fs and $V_0= 0.0005$ eV. }
\label{sfig5}
\end{figure}

\textbf{S3.3: Transition rate.}

Taking the adiabatic basis as an example, we want to discuss the transition rate and the role of decoherence time $\tau$.Here, a transient transition rate is defined as $T_{i\rightarrow j}(t)\equiv V_{ij}(t)D_{ji}(t)-D_{ij}(t)V_{ji}(t)$. This definition is obtained by expanding the commutator of equ.12. For TLS, $T_{1\rightarrow 2}(t)=\frac{2}{\hbar}\rm{Im}\left(V_{21}(t)D_{12}(t)\right)$.
Taking the approximation $\abs{V_0}\ll\hbar\omega_{12}$, $V_{21}\approx\frac{\omega V_0}{\omega_{12}} e^{-i\omega t-\eta t}$ and
\begin{align}
    D_{12}(t)= \frac{\omega V_0}{\hbar\omega_{12}} \frac{(\omega-\omega_{12})+i(\frac{1}{\tau}-\eta)}{(\omega-\omega_{12})^{2} + (\frac{1}{\tau}-\eta)^{2}}e^{i\omega_{12} t-\frac{t}{\tau}} \Delta D - \frac{\omega V_0}{\hbar\omega_{12}}\frac{(\omega-\omega_{12})+i(\frac{1}{\tau}-\eta)}{(\omega-\omega_{12})^{2} + (\frac{1}{\tau}-\eta)^{2}}e^{i\omega t - \eta t}\Delta D.
\end{align}
We have
\begin{align}
    T_{1\rightarrow 2}(t)&=\frac{2}{\hbar}\rm{Im}\left(V_{21}(t)D_{12}(t)\right)\nonumber\\
&=-2\left(\frac{\omega V_0}{\hbar\omega_{12}}\right)^2\frac{1}{(\omega-\omega_{12})^2+(\frac{1}{\tau}-\eta)^2}\rm{Im}\left\{\left[(\omega-\omega_{12})+i\left(\frac{1}{\tau}-\eta\right)\right]\left[e^{-i(\omega-\omega_{12})t-(\frac{1}{\tau}+\eta)t}-e^{-2\eta t}\right]\right\}\nonumber\\
&=-2\left(\frac{\omega V_0}{\hbar\omega_{12}}\right)^2\frac{1}{(\omega-\omega_{12})^2+(\frac{1}{\tau}-\eta)^2}\nonumber\\
&\times\left\{(\omega-\omega_{12})\rm{Im}\left(e^{-i(\omega-\omega_{12})t}\right)e^{-(\frac{1}{\tau}+\eta)t}+\left(\frac{1}{\tau}-\eta\right)\rm{Re}\left(e^{-i(\omega-\omega_{12})t}\right)e^{-(\frac{1}{\tau}+\eta)t}-\left(\frac{1}{\tau}-\eta\right)e^{-2\eta t}\right\}.
\end{align}

Here, for the small $\tau$ limit as $\frac{1}{\tau}\gg\omega-\omega_{12}$ and $\frac{1}{\tau}\gg\eta$. Then,
\begin{align}
    T_{1\rightarrow 2}(t)&\approx-2\left(\frac{\omega V_0}{\hbar\omega_{12}}\right)^2\tau^2\left[\frac{1}{\tau}\rm{Re}\left(e^{-i(\omega-\omega_{12})t}\right)e^{-\frac{t}{\tau}}-\frac{1}{\tau}e^{-2\eta t}\right]\nonumber\\
    &=-2\left(\frac{\omega V_0}{\hbar\omega_{12}}\right)^2\tau\left[\rm{Re}\left(e^{-i(\omega-\omega_{12})t}\right)e^{-\frac{t}{\tau}}-e^{-2\eta t}\right].
\end{align}
If we integrate the transition rate over a short period to find out the accumulated transitions, it yields:
$\int_0^\tau T_{1\rightarrow 2}(t)dt\approx2\left(\frac{\omega V_0}{\hbar\omega_{12}}\right)^2\tau\left[\tau-\rm{Re}\left(e^{-i(\omega-\omega_{12})t}\right)(1-e^{-1})\tau\right]\propto\tau^2$. Since $\omega-\omega_{12}$ is changing slowly, the integration over a short time will be a constant value.
This derivation is to show that even without the perturbation, decoherence itself can introduce real transitions. Particularly, when $\tau$ is extremely small ($\tau\rightarrow0$), such transition will also vanish. This is consistent to the quantum Zeno effect as well as our calculations.
Here, we only show it for the adiabatic basis case and it should be similar for the fixed basis case.

\textbf{S3.4: Multi-phonon derivation.}

Now considering the case of multiple phonons, the Hamiltonian can be written as:
\begin{align}
    \mathbf{H}(t)&=\begin{bmatrix}
\epsilon_1& \sum_{s}V_{s}e^{i\omega_{s}t-\eta t} \\
\sum_{s}V_{s}e^{-i\omega_{s}t-\eta t}&\epsilon_2
\end{bmatrix}.
\end{align}
\noindent Extending from the mono-phonon case, the non-adiabatic coupling matrix can be obtained as:
\begin{align}
    V_{21}(t)=V_{12}(t)^{*} \approx\frac{e^{-\eta t}}{\omega_{12}} \sum_{s}\omega_{s}V_{s} e^{-i\omega_{s} t}
\end{align}

\noindent Similarly, we have:
\begin{align}
    D_{22}(t=\infty)
    &=\frac{2}{\left(\hbar\omega_{12}\right)^{2}}\sum_{sk}\omega_{s}V_{s}\omega_{k}V_{k}\frac{1}{\left(\omega_{k}-\omega_{12} \right)^{2}+\left(\frac{1}{\tau}-\eta\right)^{2} }\times\nonumber\\
    &\left[
    \frac{\left(\omega_{s}-\omega_{12} \right)\left(\omega_{k}-\omega_{12} \right)-\left(\frac{1}{\tau}-\eta \right)\left(\frac{1}{\tau}+\eta \right)}{\left(\omega_{s}-\omega_{12} \right)^{2}+\left(\frac{1}{\tau}+\eta \right)^{2}}+
    \frac{\left(\omega_{k}-\omega_{12} \right)\left(\omega_{k}-\omega_{s} \right)+2\eta\left(\frac{1}{\tau}-\eta \right)}{\left(\omega_{k}-\omega_{s} \right)^{2} + 4\eta^{2}}
    \right]
\end{align}
\noindent When there are only two modes ($s=1,2$ and $k=1,2$), the above formula becomes:
\begin{align}
   &D_{22}(t=\infty)=
\nonumber\\
 &\left( \frac{ V_1\omega_1}{\hbar\omega_{12}}\right)^{2} \frac{\left(\frac{1}{\tau \eta} +1\right)}{( \omega_1-\omega_{12})^{2}+\left(\frac{1}{\tau}+\eta\right)^{2}}+\left( \frac{ V_2\omega_2}{\hbar\omega_{12}}\right)^{2} \frac{\left(\frac{1}{\tau \eta} +1\right)}{( \omega_2-\omega_{12})^{2}+\left(\frac{1}{\tau}+\eta\right)^{2}}
\nonumber\\
&+\frac{2}{(\hbar\omega_{12})^{2}}\frac{V_1\omega_1V_2\omega_2}{(\omega_2-\omega_{12})^2+(\frac{1}{\tau}-\eta)^2}\left[    \frac{\left(\omega_{1}-\omega_{12} \right)(\omega_2-\omega_{12})-\left(\frac{1}{\tau}-\eta \right)\left(\frac{1}{\tau}+\eta \right)}{\left(\omega_{1}-\omega_{12} \right)^{2}+\left(\frac{1}{\tau}+\eta \right)^{2}}-
    \frac{(\omega_2-\omega_{12})(\omega_1-\omega_2)-2(\frac{1}{\tau}-\eta)\eta}{(\omega_1-\omega_2)^2+4\eta^2}\right]
\nonumber\\
&+\frac{2}{\left(\hbar\omega_{12}\right)^{2}}\frac{V_1\omega_1V_2\omega_2}{(\omega_1-\omega_{12})^2+(\frac{1}{\tau}-\eta)^2}\left[    \frac{\left(\omega_{1}-\omega_{12} \right)(\omega_2-\omega_{12})-\left(\frac{1}{\tau}-\eta \right)\left(\frac{1}{\tau}+\eta \right)}{\left(\omega_{2}-\omega_{12} \right)^{2}+\left(\frac{1}{\tau}+\eta \right)^{2}}-
    \frac{(\omega_1-\omega_{12})(\omega_2-\omega_1)-2(\frac{1}{\tau}-\eta)\eta}{(\omega_1-\omega_2)^2+4\eta^2}\right]
\end{align}
From this formula, the first term is the contribution from he first phonon. The second term is from the second phonon. The 3rd and 4th terms are the interference terms of these two phonons.

\textbf{S4. Computational details of monolayer $\mathrm{WS_2}$}

\textbf{S4.1: First-principle calculation details.}

First-principle calculations are performed with the plane-wave package Quantum espresso (QE)\cite{giannozzi2009quantum}.
The norm-conserving (ONCV) pseudopotentials are used to describe the electron-ion interactions\cite{HamannDR2013OnVp}.
The Perdew-Burke-Ernzerhof (PBE) exchange-correlation functional is applied with the empirical DFT-D method to capture the vdW interaction\cite{PerdewJ.P.1998PBaE,GrimmeStefan2006SGdf,BaroneVincenzo2009Raet} .
An energy cutoff of 45 Ryd is used to converge the charge density.

\textbf{S4.2: Electronic band structure.}

\begin{figure}[H]
\centering
\includegraphics[width=0.8\columnwidth]{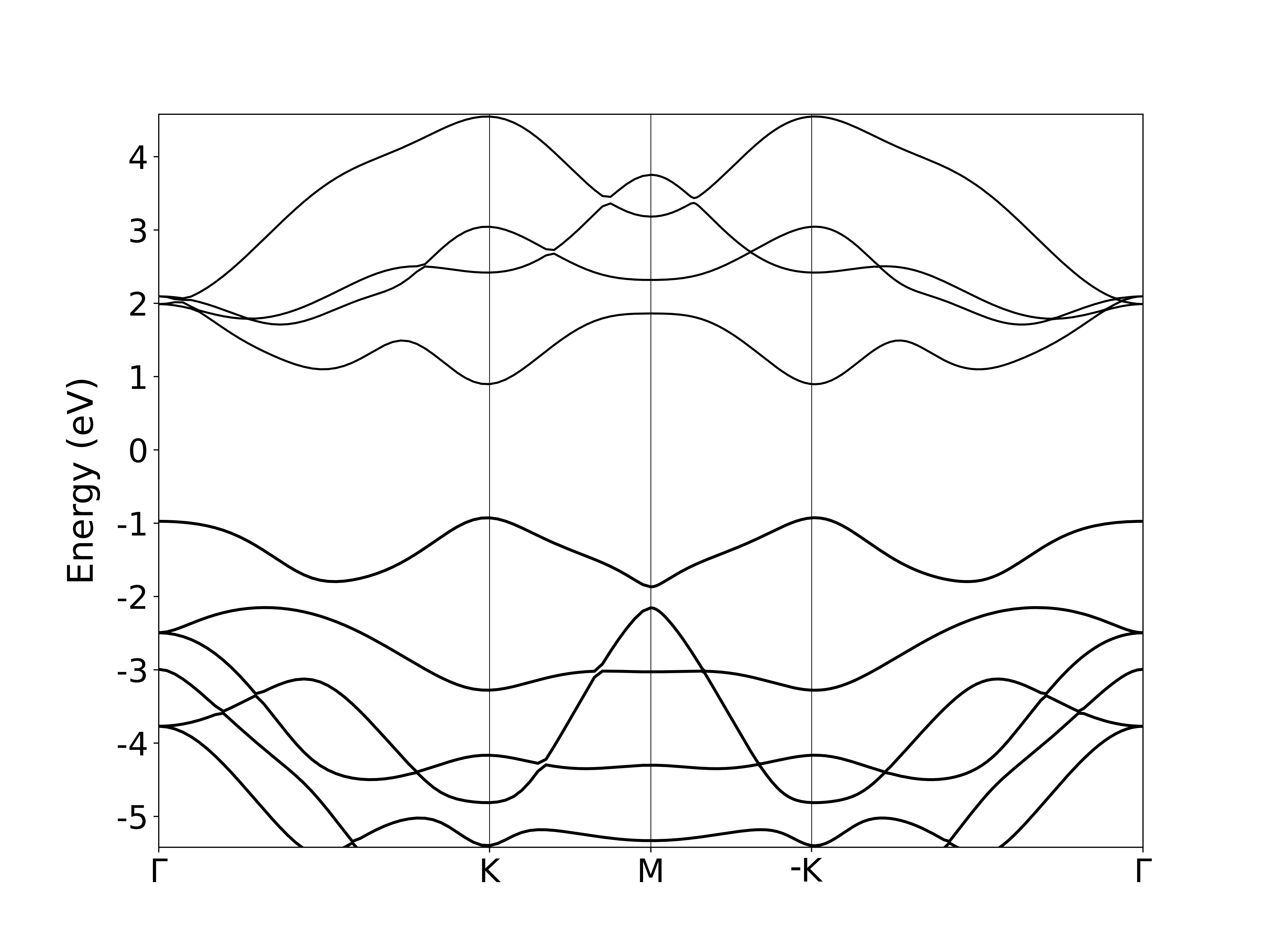}
\caption{Electronic band structure of $\mathrm{WS_2}$ unit cell. 0 of y-axis is set as the middle of the band gap.}
\label{sfig6}
\end{figure}

Shown in SI Fig.6 is the electronic band structure of $\mathrm{WS_2}$ unit cell. The band gap is 1.82 eV.

\textbf{S4.3: Supercell eigen energy time evolution.}

Shown in SI Fig.7 is the eigen energy time evolution of $\mathrm{WS_2}$ $6\times6\times1$ supercell. The three states below zero are used in this work to evolve the hole population. Two of them (VBM and VBM-1) come from the $\mathbf{k}=\left(\frac{1}{3}, \frac{1}{3},0\right)$ and $-\mathbf{k}=\left(-\frac{1}{3}, -\frac{1}{3},0\right)$ of the $\mathrm{WS_2}$ unit cell respectively, and they are degenerate in the equilibrium $\mathrm{WS_2}$. The state with a lower energy comes from the $\Gamma$ of the $\mathrm{WS_2}$ unit cell. There exist a small gap between these three states. SI Fig.8 shows the evolution of these states under $1\%$ tensile strain, stress-free, and $0.4\%$ compression. 

\begin{figure}[H]
\centering
\includegraphics[width=0.9\columnwidth]{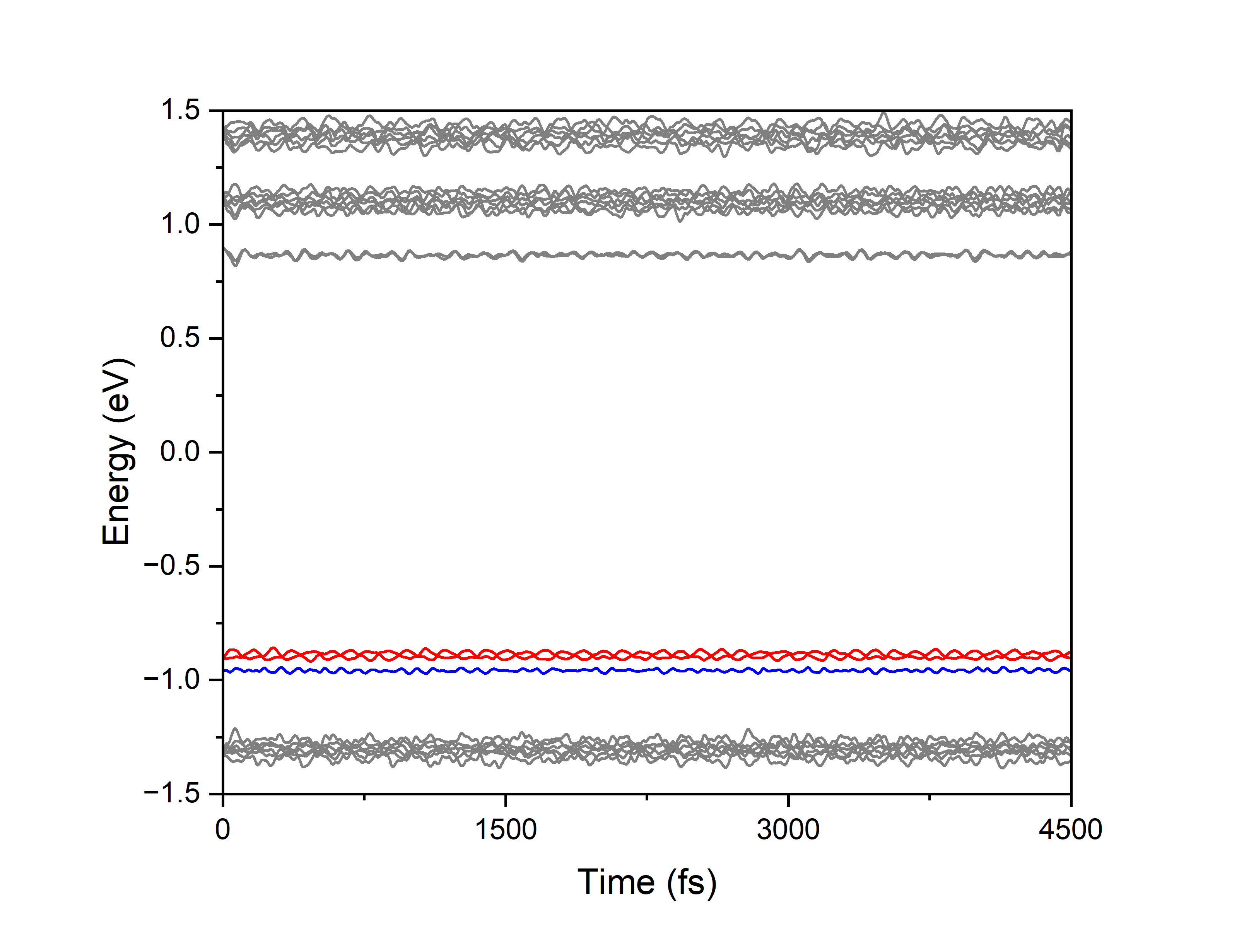}
\caption{The eigen energies time evolution of $\mathrm{WS_2}$ $6\times6\times1$ supercell. 0 of y-axis is set as the middle of the band gap at $t=0$. Red lines represent the VBM and VBM-1 states, while a blue line denotes the VBM-2 state.}
\end{figure}

\begin{figure}[H]
\centering
\includegraphics[width=1\columnwidth]{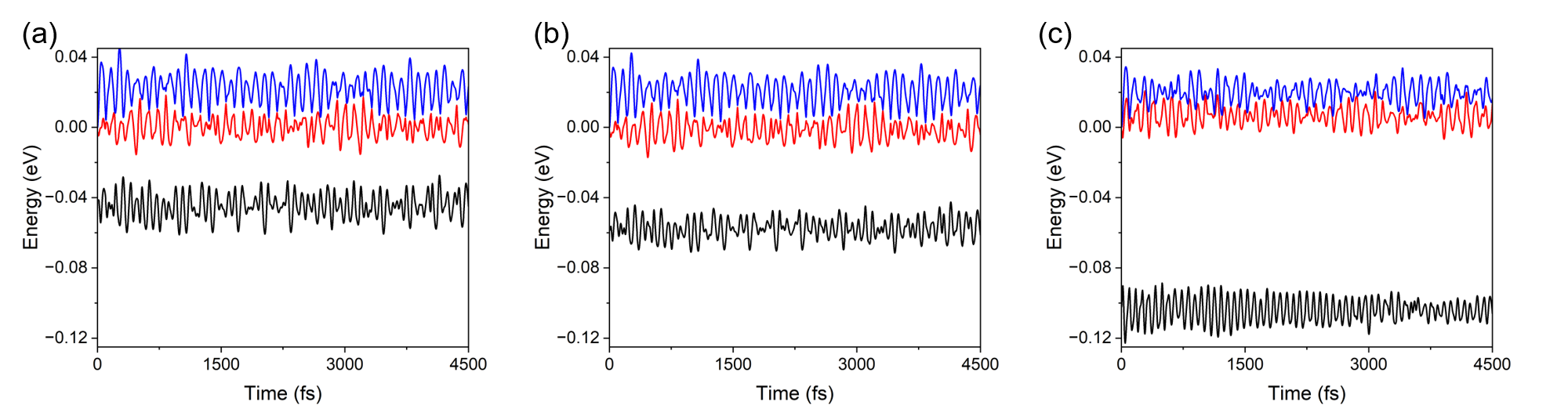}
\caption{Three states eigen energy time evolution of (a) $1\%$ tensile strain, (b) stress-free, and (c) $0.4\%$ compression $\mathrm{WS_2}$ $6\times6\times1$ supercell. 0 of y-axis is set as the VBM state energy at $t=0$. Blue line is the VBM state. Red line is the VBM-1 state. Black line is the VBM-2 state.}
\label{sfig6}
\end{figure}

\textbf{S4.4: Calculation details of molecular dynamics.}

The \textit{ab initio} ground-state molecular dynamics (AIMD) simulations are carried out using the $6\times6\times1$ supercell with a single $\Gamma$ k-point at 100 K with a time step $\Delta t$ = 2 fs.  During the AIMD, eigen energies, eigen wave functions and the overlaps between eigen states of two successive MD steps are stored for the postprocessing NAMD calculation.

\textbf{S4.5: Calculation details of density functional perturbation theory.}

The electron-phonon coupling (EPC) strength and phonon frequency are based on calculations that are performed with ONCV pseudopotentials, PBE exchange-correlation functionals and DFT-D method\cite{PerdewJ.P.1998PBaE,HamannDR2013OnVp,GrimmeStefan2006SGdf,BaroneVincenzo2009Raet}. All of the first-principles DFT calculations are performed with the plane-wave package QE\cite{giannozzi2009quantum}. A $6 \times6\times 1$ k-point grid with an energy cutoff of 45 Ryd is used to relax the atomic positions of the monolayer $\mathrm{WS_2}$ unit cell (3 atoms). Using the converged charge density, phonon calculation is performed followed by an EPC calculation. The EPC matrix element is used to construct the effective TLS. Convergence thresholds for energies and fores are set to $1\times10^{-4}$ Ryd and $1\times10^{-3}$ Ryd/Bohr.

\textbf{S5. Effective TLS.}

For a three-level system in the $\left|1\right>,\ \left|2\right> \rm and \left|3\right>$ basis, the Hamiltonian is defined as:
\begin{align}
    H(t)&=\begin{pmatrix}
0&V&V\\
V^*&\Delta&V'\\
V^*&V'^*&\Delta
\end{pmatrix},
\end{align}
where $V$ and $V'$ are functions of time $t$.

When transforming to an effective TLS Hamiltonian, first, the Hamiltonian elements related to the state $\left|2\right>$ and $\left|3\right>$ are needed to rotate to vanish the off-diagonal terms (i.e. $H_{13}$ and $H_{31}$).
Therefore, through the unitary matrix:
\begin{align}
    U=\begin{pmatrix}
1&0&0\\
0&\frac{1}{\sqrt{2}}&\frac{1}{\sqrt{2}}\\
0&\frac{1}{\sqrt{2}}&-\frac{1}{\sqrt{2}}\\
\end{pmatrix}
\end{align}
we can obtain 
\begin{align}
    H'=U^{\dagger}HU=
\begin{pmatrix}
0&\sqrt{2}V&0\\
\sqrt{2}V^*&\Delta+\mathrm{Re}V'&-i\mathrm{Im}V'\\
0&i\mathrm{Im}V'&\Delta-\mathrm{Re}V'
\end{pmatrix}.
\end{align}
Here, it is noted that if $V'$ is real, $H'$ will automatically become decoupled between $\left|1\right>$, $\left|2'\right>$ and $\left|3'\right>$.  We see that the off-diagonal term becomes $\sqrt{2}V$, which is easy to understand. The total transition rate from 1 to other states will be $\gamma\propto|V|^2+|V|^2$. Now, by combining state $\ket{2}$ and $\ket{3}$ as a single state, the effective transition rate will not change and $\gamma\propto|V_{\rm eff}|^2$, which indicates $V_{\rm eff}=\sqrt{2}V$.
Also, the standard effective Hamiltonian projection method $P\tilde{H}P=PHP+PHQ\frac{1}{E-QHQ}QHP$ ($P$ and $Q$ are projection operators) could be used and it yields:
\begin{align}
    H_{\rm{eff}}(E)=\begin{pmatrix}
0&\sqrt{2}V\\
\sqrt{2}V^*&\Delta+\mathrm{Re}V'+\frac{|\mathrm{Im}V'|^2}{E-\Delta+\mathrm{Re}V'}\\
\end{pmatrix},
\end{align}
which is a function of energy $E$ (since the perturbation will be different depending on different energy regions).
If we focus on the energy around 0 ($E=0)$, we obtain
$H_{\rm{eff}}(E)=\begin{pmatrix}
0&\sqrt{2}V\\
\sqrt{2}V^*&\Delta+\mathrm{Re}V'-\frac{|\mathrm{Im}V'|^2}{\Delta-\mathrm{Re}V'}\\
\end{pmatrix}$. For an approximation when $\abs{V'}\ll \abs{\Delta}$, we have 
\begin{align}
    H_{\rm{eff}}\approx\begin{pmatrix}
0&\sqrt{2}V\\
\sqrt{2}V^*&\Delta\\
\end{pmatrix},
\end{align}
which is same to SI equ.31.

\textbf{S6. Electron-phonon coupling for supercell states.}

If $\phi_i$ and $\phi_j$ are eigen states of supercells, we want to compute the electron-phonon coupling (EPC) between these two states using unit-cell EPC matrix element. The supercell EPC matrix element is:
\begin{align}
    V_{ij}(t)=\left<\phi_i\middle|\Delta V(t)\middle|\phi_j\right>,
\end{align}
where $\Delta V(t)\equiv V(t)-V(t=0)$ is the Kohn-Sham potential change compared to the equilibrium lattice at time $t$. Then 
\begin{align}
    V_{ij}(t)=\sum_k\left<\phi_i\middle|\left.\frac{\partial V}{\partial \mathbf{R}_k}\right|_{\mathbf{R}_0}\mathbf{\Delta R}_k(t)\middle|\phi_j\right>
\end{align}
where $\mathbf{R}_{k}$ is the position of atom $k$. This inner product is integrated over the whole space. However, if the supercell is computed with the single-k point, the integration will be performed within the supercell only (following the Born-von Karman boundary condition (BvK)). 

The atomic displacement can be expanded as:
\begin{align}
    \mathbf{\Delta R}_{\kappa,p}(t)=\frac{1}{\sqrt{N_p}}\sum_{\mathbf{q},\nu}\mathbf{e}_{\kappa\nu}(\mathbf{q})e^{i\mathbf{q}\cdot\mathbf{r}}\frac{Q_{\mathbf{q}\nu}(t)}{\sqrt{m_\kappa}}
\end{align}
where $\kappa$ and $p$ indicate the atom at basis-$\kappa$ at unit cell-$p$. $N_p$ is the number of unit cells under the BvK periodic boundary condition. 
$Q_{\mathbf{q}\nu}(t)$ is the normal-mode coordinate for mode $\nu$ at wavevector $\mathbf{q}$. $\mathbf{e}_{\kappa\nu}(\mathbf{q})$ is the eigen vector of the dynamical matrix and it satisfies orthonormalization between different atom and different modes.

Plug in the expansion back to the EPC expression, we have
\begin{align}
   V_{ij}(t)=\frac{1}{\sqrt{N_p}}\left<\phi_i\middle|\sum_{\kappa,p}\left.\frac{\partial V}{\partial \mathbf{R}_{\kappa,p}}\right|_{\mathbf{R}_0}\sum_{\mathbf{q},\nu}\mathbf{e}_{\kappa\nu}(\mathbf{q})e^{i\mathbf{q}\cdot\mathbf{R}_p}\frac{Q_{\mathbf{q}\nu}(t)}{\sqrt{m_\kappa}}\middle|\phi_j\right>.
\end{align}
Here, the sum over k is replaced by $\kappa$ and $p$.

Now, a band unfolding of the supercell eigen states to the unit cell eigen states is performed. We have
\begin{align}\phi_{\mathbf{K}}=\sum_{m\mathbf{k}}\alpha_{m\mathbf{k}}\widetilde{\phi}_{m\mathbf{k}}
\end{align}
where $\widetilde{\phi}_{m\mathbf{k}}$ is the unit cell eigen state. $\alpha$ is the projection coefficient. $\mathbf{k}$ is the wavevector in the unit cell’s Brillouin zone that must satisfy $\mathbf{k}=\mathbf{K}+\mathbf{G}$ where $\mathbf{K}$ is the wavevector in the supercell’s Brillouin zone and $\mathbf{G}$ is the reciprocal lattice vector of the supercell. Here, we only consider trivial band-unfolding for $\rm WS_2$ case, i.e. the supercell structure is simply the repetition of unit cells. For example, $\phi_i$ is the VBM-2 of the supercell:
\begin{align}
    \phi_i=\frac{1}{\sqrt{N_p}}e^{i\mathbf{k}\cdot\mathbf{r}}\widetilde{u}_{m\mathbf{k}}
\end{align}
and $\phi_j$ is the VBM-1 or VBM of the supercell:
\begin{align}
    \phi_j=\frac{1}{\sqrt{2}}\frac{1}{\sqrt{N_p}}e^{i\mathbf{k'}\cdot\mathbf{r}}\widetilde{u}_{n,\mathbf{k'}}+\frac{1}{\sqrt{2}}\frac{1}{\sqrt{N_p}}e^{-i\mathbf{k'}\cdot\mathbf{r}}\widetilde{u}_{n,\mathbf{-k'}}. 
\end{align}
Here, $\mathbf{k}=\Gamma$ and $\mathbf{k'}=\left(\frac{1}{3},\frac{1}{3},0\right)$. For supercell, $\psi_{n\mathbf{k'}}$ and $\psi_{n,-\mathbf{k'}}$ are degenerate, therefore, their linear combinations will contribute to $\phi_j$. In our calculation, we simply find half-and-half contributions from $\mathbf{k'}$ and $-\mathbf{k'}$ for $\phi_j$; but the following derivation could be applied to any trivial band unfolding cases.
\begin{align}
    V_{ij}(t)&=\frac{1}{\sqrt{2}}\frac{1}{\sqrt{N_p}}\frac{1}{N_p}\sum_{\mathbf{q},\nu}\left<\widetilde{u}_{m\mathbf{k}}e^{-i\mathbf{k}\cdot\mathbf{r}}\middle|\sum_{\kappa,p}\left.\frac{\partial V}{\partial \mathbf{R}_{\kappa,p}}\right|_{\mathbf{R}_0}
    \frac{\mathbf{e}_{\kappa\nu}(\mathbf{q})}{\sqrt{M_\kappa}}e^{i\mathbf{q}\mathbf{R}_p}
\middle|\widetilde{u}_{n\mathbf{k'}}e^{i\mathbf{k'}\cdot\mathbf{r}}\right>Q_{\mathbf{q}\nu}(t)\nonumber\\
    &+\frac{1}{\sqrt{2}}\frac{1}{\sqrt{N_p}}\frac{1}{N_p}\sum_{\mathbf{q},\nu}\left<\widetilde{u}_{m\mathbf{k}}e^{-i\mathbf{k}\cdot\mathbf{r}}\middle|\sum_{\kappa,p}\left.\frac{\partial V}{\partial \mathbf{R}_{\kappa,p}}\right|_{\mathbf{R}_0}
    \frac{\mathbf{e}_{\kappa\nu}(\mathbf{q})}{\sqrt{M_\kappa}}e^{i\mathbf{q}\mathbf{R}_p}
\middle|\widetilde{u}_{n,-\mathbf{k'}}e^{-i\mathbf{k'}\cdot\mathbf{r}}\right>Q_{-\mathbf{q}\nu}(t)\nonumber\\
    &=\frac{1}{\sqrt{2}}\frac{1}{\sqrt{N_p}}\frac{1}{N_p}\sum_{\mathbf{q},\nu}\left<\widetilde{u}_{m\mathbf{k}}\middle|\sum_{\kappa,p}\left.\frac{\partial V}{\partial \mathbf{R}_{\kappa,p}}\right|_{\mathbf{R}_0}
 \frac{\mathbf{e}_{\kappa\nu}(\mathbf{q})}{\sqrt{M_\kappa}}
 e^{-i\mathbf{q}\left(\mathbf{r}-\mathbf{R}_p\right)}e^{i\left(\mathbf{q}+\mathbf{k}+\mathbf{k}'\right)\cdot\mathbf{r}}\middle|\widetilde{u}_{n\mathbf{k'}}\right>Q_{\mathbf{q}\nu}(t)\nonumber\\
    &+\frac{1}{\sqrt{2}}\frac{1}{\sqrt{N_p}}\frac{1}{N_p}\sum_{\mathbf{q},\nu}\left<\widetilde{u}_{m\mathbf{k}}\middle|\sum_{\kappa,p}\left.\frac{\partial V}{\partial \mathbf{R}_{\kappa,p}}\right|_{\mathbf{R}_0}
 \frac{\mathbf{e}_{\kappa\nu}(\mathbf{q})}{\sqrt{M_\kappa}}
 e^{-i\mathbf{q}\left(\mathbf{r}-\mathbf{R}_p\right)}e^{i\left(\mathbf{q}+\mathbf{k}-\mathbf{k}'\right)\cdot\mathbf{r}}\middle|\widetilde{u}_{n,-\mathbf{k'}}\right>Q_{-\mathbf{q}\nu}(t).
\end{align}
It can be demonstrated that $\sum_{\kappa,p}\left.\frac{\partial V}{\partial \mathbf{R}_{\kappa,p}}\right|_{\mathbf{R}_0}\frac{\mathbf{e}_{\kappa\nu}(\mathbf{q})}{\sqrt{M_\kappa}}e^{-i\mathbf{q}\cdot \left(\mathbf{r}-\mathbf{R}_p\right)} $ is periodic for unit cell lattice vector $\mathbf{R}$. Then the integral will vanish unless $\mathbf{q}=-\mathbf{k'}-\mathbf{k}$ (which is $\mathbf{q}=(-1/3,-1/3,0)$ (or its negative)). Following that, the integral over the whole space can be reduced to integration within the unit cell.
\begin{align}
V_{ij}(t)&=\frac{1}{\sqrt{2}}\frac{1}{\sqrt{N_p}}\sum_{\nu}\left<\widetilde{u}_{m\mathbf{k}}\middle|\sum_{\kappa,p}\left.\frac{\partial V}{\partial \mathbf{R}_{\kappa,p}}\right|_{\mathbf{R}_0}\frac{\mathbf{e}_{\kappa\nu}(\mathbf{q})}{\sqrt{M}_\kappa}e^{-i\mathbf{q}\cdot \left(\mathbf{r}-\mathbf{R}_p\right)}\middle|\widetilde{u}_{n\mathbf{k'}}\right>_{\rm{uc}}Q_{\mathbf{q}\nu}(t)\nonumber\\
&+\frac{1}{\sqrt{2}}\frac{1}{\sqrt{N_p}}\sum_{\nu}\left<\widetilde{u}_{m\mathbf{k}}\middle|\sum_{\kappa,p}\left.\frac{\partial V}{\partial \mathbf{R}_{\kappa,p}}\right|_{\mathbf{R}_0}\frac{\mathbf{e}_{\kappa\nu}(\mathbf{q})}{\sqrt{M}_\kappa}e^{-i\mathbf{q}\cdot \left(\mathbf{r}-\mathbf{R}_p\right)}\middle|\widetilde{u}_{n,-\mathbf{k'}}\right>_{\rm{uc}}Q_{-\mathbf{q}\nu}(t) \nonumber\\
&=\frac{1}{\sqrt{2}}\frac{1}{\sqrt{N_p}}\sum_{\nu}\left<\widetilde{u}_{m\mathbf{k}}\middle|\sum_{\kappa,p}\left.\frac{\partial V}{\partial \mathbf{R}_{\kappa,p}}\right|_{\mathbf{R}_0}\sqrt{\frac{\hbar}{2M_\kappa\omega_{\mathbf{q}\nu}}}\mathbf{e}_{\kappa\nu}(\mathbf{q})e^{-i\mathbf{q}\cdot \left(\mathbf{r}-\mathbf{R}_p\right)}\middle|\widetilde{u}_{n\mathbf{k'}}\right>_{\rm{uc}}Q_{\mathbf{q}\nu}(t)\sqrt{\frac{2\omega_{\mathbf{q}\nu}}{\hbar}}\nonumber\\
&+\frac{1}{\sqrt{2}}\frac{1}{\sqrt{N_p}}\sum_{\nu}\left<\widetilde{u}_{m\mathbf{k}}\middle|\sum_{\kappa,p}\left.\frac{\partial V}{\partial \mathbf{R}_{\kappa,p}}\right|_{\mathbf{R}_0}\sqrt{\frac{\hbar}{2M_\kappa\omega_{-\mathbf{q}\nu}}}\mathbf{e}_{\kappa\nu}(\mathbf{q})e^{-i\mathbf{q}\cdot \left(\mathbf{r}-\mathbf{R}_p\right)}\middle|\widetilde{u}_{n,-\mathbf{k'}}\right>_{\rm{uc}}Q_{-\mathbf{q}\nu}(t)\sqrt{\frac{2\omega_{-\mathbf{q}\nu}}{\hbar}} \nonumber\\
&\equiv \frac{1}{\sqrt{2}}\frac{1}{\sqrt{N_p}}\sum_{\nu}g_{mn,\nu}(\mathbf{\Gamma,q})Q_{\mathbf{q}\nu}(t)\sqrt{\frac{2\omega_{\mathbf{q}\nu}}{\hbar}}+\frac{1}{\sqrt{2}}\frac{1}{\sqrt{N_p}}\sum_{\nu}g_{mn,\nu}(\mathbf{\Gamma,-q})Q_{\mathbf{-q}\nu}(t)\sqrt{\frac{2\omega_{\mathbf{-q}\nu}}{\hbar}}
\end{align}

The last line is to use the definition of the first-order EPC matrix element,\cite{Giustino17p015003} where the quantity evaluated within the integral is simply the unit-cell EPC matrix element: $g_{mn,\nu}(\mathbf{q})$. Therefore, by performing a time-average over $V_{ij}(t)$ to obtain an averaged coupling, we have:
\begin{align}
    \left<\left|V_{ij}(t)\right|^2\right>&=\frac{1}{2}\frac{1}{N_p}\left<\left|\sum_{\nu}g_{mn,\nu}(\mathbf{\Gamma,q})Q_{\mathbf{q}\nu}(t)\sqrt{\frac{2\omega_{\mathbf{q}\nu}}{\hbar}}+\sum_{\nu}g_{mn,\nu}(\mathbf{\Gamma,-q})Q_{\mathbf{-q}\nu}(t)\sqrt{\frac{2\omega_{\mathbf{-q}\nu}}{\hbar}}\right|^2\right>\nonumber\\
&=\frac{1}{2}\frac{1}{N_p}\sum_{\nu}\left|g_{mn,\nu}(\mathbf{\Gamma,q})\right|^2\left<Q_{\mathbf{q}\nu}^*(t)Q_{\mathbf{q}\nu}(t)\right>\frac{2\omega_{\mathbf{q}\nu}}{\hbar}+\sum_{\nu}\left|g_{mn,\nu}(\mathbf{\Gamma,-q})\right|^2\left<Q_{\mathbf{-q}\nu}^*(t)Q_{\mathbf{-q}\nu}(t)\right>\frac{2\omega_{\mathbf{-q}\nu}}{\hbar}\nonumber\\
&=\frac{1}{N_p}\sum_\nu\left|g_{mn,\nu}(\mathbf{\Gamma,q})\right|^2\frac{\hbar}{\omega_{\mathbf{q}\nu}}(n_{\mathbf{q}\nu}+\frac{1}{2})\frac{2\omega_{\mathbf{q}\nu}}{\hbar}\nonumber\\
&\approx \frac{1}{N_p}\sum_\nu\left|g_{mn,\nu}(\mathbf{\Gamma,q})\right|^2\frac{2k_{\rm B}T}{\hbar\omega_{\mathbf{q}\nu}}.
\end{align}
The second line is obtain by expanding the $\abs{\cdots}^2$ term and using the orthogonality of normal-mode coordinates under time-average (orthogonal between different $\mathbf{q}$ and $\nu$ under harmonic approximation). The third line is to use the relation of normal-mode coordinates $\left<Q_{\mathbf{q}\nu}^*(t)Q_{\mathbf{q}\nu}(t)\right>=\frac{\hbar}{\omega_{\mathbf{q}\nu}}(n_{\mathbf{q}\nu}+\frac{1}{2})\approx\frac{k_{\rm B}T}{\omega_{\mathbf{q}\nu}^2}$ where $n$ is the number of phonons. The last line approximation is to use the high-temperature limit.

Here, note that owing to the symmetry, $\omega_{\mathbf{q}\nu}=\omega_{-\mathbf{q}\nu}$, $g_{mn,\nu}(\mathbf{\Gamma},\mathbf{q})=g_{mn,\nu}^*(\mathbf{\Gamma},\mathbf{-q})$. Now, we want to obtain the averaged effective contribution from each mode $\nu$ to $V_{ij}$. Then, we can define:
\begin{align}
    \overline{V}_{ij,\nu}=\frac{1}{\sqrt{N_p}}\left|g_{mn,\nu}(\mathbf{\Gamma},\mathbf{q})\right|\sqrt{\frac{2k_{\rm B}T}{\hbar\omega_{\mathbf{q}\nu}}}.
\end{align}

\textbf{S7. One phonon results of $\mathrm{WS_2}$ and  effective TLS.}

Based on equ.33 and equ.44, we can map the parameters of the real system onto an effective TLS and use the P-matrix NAMD simulations to determine the respective hole population in both the $\mathrm{WS_2}$ system and the effective TLS. By using only one phonon mode, the results of populations for TLS and DFT are inconsistent, in particular for the small-gap structures (SI Fig. 9(a) and (b)).

\begin{figure}
\centering
\includegraphics[width=1\columnwidth]{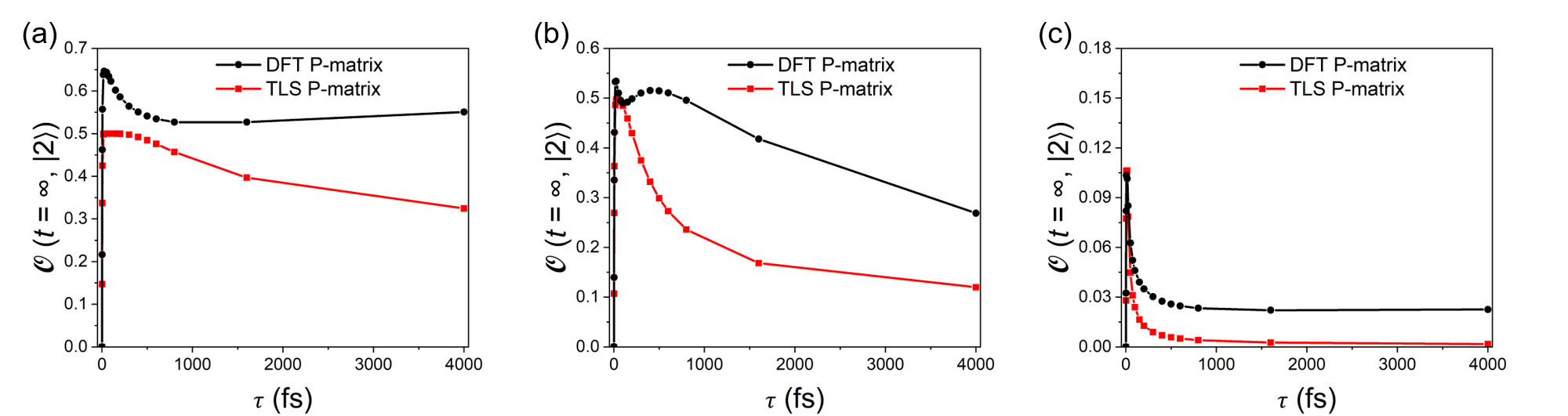}
\caption{The NAMD-simulated excited carrier of the real system and the TLS. Only one phonon is considered with the phonon energy is 0.046 eV and EPC strength is 0.007 eV. (a) Tensiled structure ($1\%$ tensile): energy gap is 0.056 eV. (b) Stress-free structure: energy gap is 0.067 eV. c) compressed structure ($0.4\%$ compression): energy gap is 0.12 eV. The initial state is a fully occupied state on VBM-2.}
\label{sfig6}
\end{figure}

\bibliography{SI}